\documentclass[twocolumn,trackchanges]{aastex701}
\usepackage{multirow}
\usepackage{amsmath}

\begin{document}

\title{Emission spectroscopy of KELT-2A b with Gemini-North/IGRINS-2: {hints of a stellar C/O ratio and sub-stellar metallicity}}

\author[orcid=0000-0001-8877-0242,sname='Hayashi']{Yuya Hayashi}
\affiliation{Department of Multi-Disciplinary Sciences, Graduate School of Arts and Sciences, The University of Tokyo, 3-8-1 Komaba, Meguro, Tokyo 153-8902, Japan}
\email[show]{yuya-hayashi@g.ecc.u-tokyo.ac.jp}

\author[orcid=0000-0003-4698-6285,sname='Nugroho']{Stevanus K. Nugroho}
\affiliation{Department of Earth and Planetary Sciences, Institute of Science Tokyo, 2-12-1 Ookayama, Meguro, Tokyo 152-8551, Japan}
\affiliation{Astrobiology Center, 2-21-1 Osawa, Mitaka, Tokyo 181-8588, Japan}
\affiliation{National Astronomical Observatory of Japan, 2-21-1 Osawa, Mitaka, Tokyo 181-8588, Japan}
\email{skristiantonugroho@gmail.com}

\author[orcid=0000-0003-2493-912X,sname='Notsu']{Shota Notsu}
\affiliation{Department of Earth and Planetary Science, Graduate School of Science, The University of Tokyo, 7-3-1 Hongo, Bunkyo-ku, Tokyo 113-0033, Japan}
\affiliation{Star and Planet Formation Laboratory, Pioneering Research Institute, RIKEN, 2-1 Hirosawa, Wako, Saitama 351-0198, Japan}
\email{shota.notsu@eps.s.u-tokyo.ac.jp}

\author[orcid=0000-0001-6516-4493,sname='Changeat']{Quentin Changeat}
\affiliation{Kapteyn Institute, University of Groningen, 9747 AD Groningen, NL}
\email{q.changeat@rug.nl}

\author[0000-0002-9946-5259]{Peter C. B. Smith}
\affiliation{School of Earth and Space Exploration, Arizona State University, 781 Terrace Mall, Tempe, AZ, 85287, USA}
\email{psmith57@asu.edu}

\author[0009-0005-4890-3326]{Krishna Kanumalla}
\affiliation{School of Earth and Space Exploration, Arizona State University, 781 Terrace Mall, Tempe, AZ, 85287, USA}
\email{skanuma1@asu.edu}

\author[0000-0002-2338-476X]{Michael R. Line}
\affiliation{School of Earth and Space Exploration, Arizona State University, 781 Terrace Mall, Tempe, AZ, 85287, USA}
\email{mrline@asu.edu}

\author[0000-0002-9142-6378]{Jorge A. Sanchez}
\affiliation{School of Earth and Space Exploration, Arizona State University, 781 Terrace Mall, Tempe, AZ, 85287, USA}
\email{jasanchez@asu.edu}

\author[orcid=0000-0001-8002-8473,sname='Yoshida']{Tomohiro C. Yoshida}
\affiliation{Dipartimento di Fisica, Universit\`{a} degli Studi di Milano, Via Celoria 16, 20133 Milano, Italy}
\affiliation{National Astronomical Observatory of Japan, 2-21-1 Osawa, Mitaka, Tokyo 181-8588, Japan}
\email{tomohiroyoshida.astro@gmail.com}

\author[orcid=0000-0003-3290-6758,sname='Ohno']{Kazumasa Ohno}
\affiliation{National Astronomical Observatory of Japan, 2-21-1 Osawa, Mitaka, Tokyo 181-8588, Japan}
\email{ohno.k.ab.715@gmail.com}

\author[orcid=0000-0002-7058-7682,sname='Nomura']{Hideko Nomura}
\affiliation{National Astronomical Observatory of Japan, 2-21-1 Osawa, Mitaka, Tokyo 181-8588, Japan}
\affiliation{Department of Astronomical Science, The Graduate University for Advanced Studies, SOKENDAI, 2-21-1 Osawa, Mitaka, Tokyo 181-8588, Japan}
\email{hideko.nomura@nao.ac.jp}

\author[orcid=0000-0002-5331-6637,sname='Kagetani']{Taiki Kagetani}
\affiliation{Department of Multi-Disciplinary Sciences, Graduate School of Arts and Sciences, The University of Tokyo, 3-8-1 Komaba, Meguro, Tokyo 153-8902, Japan}
\affiliation{National Astronomical Observatory of Japan, 2-21-1 Osawa, Mitaka, Tokyo 181-8588, Japan}
\email{kagetani@g.ecc.u-tokyo.ac.jp}

\author[orcid=0000-0001-8511-2981,sname='Narita']{Norio Narita}
\affiliation{Astrobiology Center, 2-21-1 Osawa, Mitaka, Tokyo 181-8588, Japan}
\affiliation{Komaba Institute for Science, The University of Tokyo, 3-8-1 Komaba, Meguro, Tokyo 153-8902, Japan}
\affiliation{Instituto de Astrof\'isica de Canarias (IAC), E-38200 La Laguna, Tenerife, Spain}
\email{narita@g.ecc.u-tokyo.ac.jp}

\author[orcid=0000-0003-1205-5108,sname='Kawauchi']{Kiyoe Kawauchi}
\affiliation{Department of Physical Sciences, Ritsumeikan University, Kusatsu, Shiga 525-8577, Japan}
\email{kawauchi@fc.ritsumei.ac.jp}

\author[orcid=0000-0002-4909-5763,sname='Fukui']{Akihiko Fukui}
\affiliation{Komaba Institute for Science, The University of Tokyo, 3-8-1 Komaba, Meguro, Tokyo 153-8902, Japan}
\affiliation{Instituto de Astrof\'isica de Canarias (IAC), E-38200 La Laguna, Tenerife, Spain}
\email{afukui@g.ecc.u-tokyo.ac.jp}

\author[orcid=0000-0003-0987-1593,sname='Palle']{Enric Palle}
\affiliation{Instituto de Astrof\'isica de Canarias (IAC), E-38200 La Laguna, Tenerife, Spain}
\affiliation{Departamento de Astrof\'isica, Universidad de La Laguna (ULL), E-38206 La Laguna, Tenerife, Spain}
\email{epalle@iac.es}

\begin{abstract}
We present the results of high-resolution emission spectroscopy of the hot Jupiter KELT-2A~b using IGRINS-2 on the Gemini-North telescope, covering the $H$- and $K$-bands at a spectral resolution of $R \sim 45{,}000$. Observations were carried out over five nights, covering both pre- and post-eclipse orbital phases. Stellar and telluric contaminations are removed via singular value decomposition. We detect $\rm H_2O$ at a signal-to-noise ratio of {6.57}, confirming the previous result at a higher significance. While $\rm ^{12}CO$ is identified in the log-likelihood map near the expected planetary velocity, a definitive detection of $\rm ^{12}CO$ remains elusive given the current data quality. We perform atmospheric retrieval both with IGRINS-2 data alone and jointly with archival \textit{Spitzer}/IRAC photometric data points. The retrieved temperature–pressure profile is non-inverted, consistent with prior observations. {From the joint fit we constrain the water abundance to} $\rm log(H_2O) = {-3.78^{+0.27}_{-0.20}}$ {with the free chemistry model}. {The chemical equilibrium model yields} $\rm C/O={ 0.19^{+0.12}_{-0.07}}$, $\rm [C/H]=-0.81^{+0.41}_{-0.34}$, and $\rm [O/H] = -0.37^{+0.27}_{-0.17}${: the C/O ratio agrees with the stellar value, while the carbon and oxygen abundances are moderately depleted relative to the host star, though still consistent at the $\sim2\sigma$ level.} If confirmed, {this combination of a stellar C/O ratio with a sub-stellar metallicity is difficult to reconcile with gas accretion between the $\rm{H_2O}$ and CO snowlines, which predicts a super-stellar C/O. It instead points to the accretion of metal-depleted gas outside the CO snowline with only limited planetesimal enrichment, and hence to a disk-free, high-eccentricity migration history.}
\end{abstract}



\section{Introduction} 
The protoplanetary disk is where planets form, and its temperature varies with distance from the host star. Since each volatile species has a distinct condensation temperature, the chemical composition of the disk gas differs depending on radial distance. Therefore, under the simplest assumptions, the atmospheric composition of a gas giant can be used as a proxy for the planet's formation location \citep{Oberg_2011}. In particular, since the C/O ratio and metallicity have a significant influence on the observable atmospheric spectra of planets through the abundances of atmospheric chemical species \citep{Madhusudhan_2012}, the importance of these measurements is widely recognized (e.g., \citealt{Madhusudhan_2014}). In addition, processes such as disk chemical evolution (e.g., \citealt{Eistrup_2016, Notsu_2020}), planetesimal accretion (e.g., \citealt{Mordasini_2016, Cridland_2019}), and pebble drift and sublimation (e.g., \citealt{Booth_2017, Booth_2019, Schneider_Bitsch_2021, Ohno_2026}) predict a wider range of atmospheric compositions than the simple disk model.\par
Most atmospheric observations have utilized space-based telescopes such as Hubble Space Telescope (HST) and \textit{Spitzer}, measuring wavelength-dependent transit and eclipse depths at low-to-medium resolution ($R \sim 30$--1000) to constrain C/O ratios and metallicity (e.g., \citealt{Sing_2016, Changeat_2022, Edwards_2023}). James Webb Space Telescope (JWST) has recently transformed the field with enhanced wavelength coverage and precision, enabling significantly improved measurements of these abundance ratios (e.g., \citealt{Rustamkulov_2023, Coulombe_2023, Fu_2025}). JWST achieves a maximum resolving power of $R\sim3000$, but for transiting planets, most individual spectral lines remain unresolved.\par
High-resolution cross-correlation spectroscopy (HRCCS) with ground-based high-resolution echelle spectrographs ($R > 25{,}000$) offers a complementary approach by exploiting the planet's orbital Doppler shift to separate planetary signals from telluric and stellar contamination (e.g., \citealt{Snellen_2010, Brogi_2012}). By resolving individual spectral lines, HRCCS offers unique capabilities, including the detection of atomic metals and ionized species \citep[e.g.,][]{Hoeijmakers_2018}, and the unambiguous identification of individual molecular species \citep[e.g.,][]{Birkby_2013, Nugroho_2021}.\par
Recent developments in atmospheric retrieval with HRCCS have advanced the field beyond mere molecular detection \citep{Brogi_Line_2019, Gibson_2020}. This framework allows quantitative Bayesian constraints on model parameters through direct comparison of forward models with data — analogous to what has been routinely done in low-resolution spectroscopy. It also allows us to calculate the posteriors of model parameters taking into account the correlations between them. The HRCCS retrieval approach also enables joint analyses of space-based photometric or low-resolution spectroscopic data with ground-based high-resolution spectroscopic data, leveraging the complementary strengths of each to place more precise constraints on exoplanet atmospheres \citep[e.g.,][]{Brogi_Line_2019, Yan_2022, Smith2024_wasp77Ab}. Despite its promise, this joint approach has so far been applied to only a handful of planetary atmospheres. \par
KELT-2A~b is a transiting hot Jupiter discovered by the Kilodegree Extremely Little Telescope (KELT) Survey, orbiting a slightly evolved late F-type star with a period of 4.11 days. The planet has an equilibrium temperature of $T_{\rm eq} \sim 1700$~K, a mass of $1.52~M_{\rm J}$, and a radius of $1.29~R_{\rm J}$ \citep{Beatty_2012}. Key properties of the host star and planet are summarized in Table~\ref{tab:planet_parameters}. 
It is among the earliest planets to have been studied through a joint analysis of ground-based high-resolution spectroscopy and space-based photometry. \citet{Piskorz_2018} reported the tentative evidence of $\rm H_2O$ absorption in the \textit{L}-band using Keck/NIRSPEC, along with hints of super-solar metallicity ($\log z = 1.06^{+0.67}_{-1.17}$, where $z$ is the metallicity relative to solar) and a solar C/O ratio (${\rm C/O} = 0.53^{+0.20}_{-0.19}$) through a joint fit with \textit{Spitzer}/IRAC photometry, though these results were limited by the data quality of Keck/NIRSPEC at the time.\par
In this paper, we present emission spectroscopy of KELT-2A~b obtained with the Immersion GRating INfrared Spectrometer-2 (IGRINS-2) on the Frederick C. Gillett Gemini Telescope (Gemini-North Telescope). IGRINS-2 is a sibling instrument to IGRINS, previously mounted on Gemini-South, and provides simultaneous high-resolution ($R \sim 45{,}000$) spectroscopy across the $H$- and $K$-bands (1.45--2.5~$\mu$m) \citep{Lee_2022, Oh_2024}. This wavelength range allows us to capture the absorption features of H$_2$O and CO, the primary carbon and oxygen carriers at the temperature regime of KELT-2A~b.\par
The organization of this paper is as follows. In Section~\ref{section:Observation_Reduction}, we describe our observations and data 
reduction. Section~\ref{sec:Atmospheric_Retireval} details the construction of model spectra and the retrieval 
analysis framework. Section~\ref{sec:results} presents the results of the retrieval analysis. Section~\ref{sec:validation} shows the validation and robustness check of the planetary signals presented in Section~\ref{sec:results}.
In Section~\ref{sec:discussion}, we discuss the implication for planet formation based on the retrieved atmospheric composition. Finally, Section~\ref{sec:summary} summarizes our findings.\par
\begin{deluxetable*}{lcc}
\tablewidth{0pt}
\tablecaption{Properties for KELT-2A System\label{tab:planet_parameters}}
\tablehead{
\colhead{Property} & \colhead{Value} & \colhead{Reference}}
\startdata
\multicolumn{3}{c}{KELT-2A} \\
Mass, $M_s$ & $1.308\pm{0.028}\,M_\odot$ & (1) \\
Radius, $R_s$ & $1.828\pm{0.070}\,R_\odot$ & (1) \\
Effective Temp., $T_{\rm eff}$ & $6150\pm{50}\,{\rm K}$ & (1) \\
Carbon abundance, $\rm [C/H]$ & $0.01\pm0.09$ dex & (2) \\
Oxygen abundance, $\rm [O/H]$ & $0.35\pm0.09$ dex & (2) \\
Systemic velocity, $v_{\rm sys}$ & $-47.2\pm{0.3}\,{\rm km\,s^{-1}}$ & (3) \\
$K$-band magnitude, $K_{\rm mag}$ & $7.35\pm{0.03}$ & (4) \\
\multicolumn{3}{c}{KELT-2A\ b} \\
Period, $P$ & 4.114 days & (5) \\
Mass, $M_p$ & $1.522^{+0.078}_{-0.077}\,M_{\rm J}$ & (1) \\
Radius, $R_p$ & $1.29^{+0.065}_{-0.047}\,R_{\rm J}$ & (1) \\
Equilibrium Temp., $T_{\rm eq}$ & $1710\pm40$ K & Derived \\
Semi-major axis, $a$ & $0.05496^{+0.00039}_{-0.00036}$ au & (1) \\
Inclination, $i$ & ${88.6^\circ}^{+1.0^\circ}_{-1.4^\circ}$ & (1) \\
Transit center, $T_c$ & 2459467.1990 days & (5) \\
Eccentricity, $e$ & 0 & (1) \\
Orbital velocity, $K_{\rm p}$ & $145\pm1\,{\rm km\,s^{-1}}$ & Derived \\ 
\enddata
\tablerefs{(1) \cite{Beatty_2012}; (2) \cite{Behmard_2023}; (3) \cite{Gaia_2018}; (4) \cite{Cutri_2003}; (5) \cite{Kokori_2023}}
\end{deluxetable*}

\section{Observations and Data Reduction} 
\label{section:Observation_Reduction}
\subsection{Observation}
\label{subsec:observation}
We observed the thermal emission of KELT-2A~b using IGRINS-2 on Gemini North under Gemini
program ID GN-2025A-Q-212 (PI: Y. Hayashi) and GN-2025B-Q-209 (PI: Y. Hayashi).  Data were collected over five nights from February 2025 to January 2026, covering both pre- and post-eclipse phases. For each night, we repeatedly took exposures adopting an AB-BA nodding pattern, with integration times of 91.45 s for the three nights in 2025 and 120 s for the two nights in 2026. Hereafter, we refer to each AB-pair as a frame. KELT-2A has a companion star, KELT-2B, located at a separation of $2".3$ almost due north. To prevent contamination from the companion star, we fixed the slit position angle at $30^\circ$ during observations. Slit viewer camera images confirmed that the companion star was well spatially separated from the target on each night. The observation dates and times, orbital phases, number of frames, and typical signal-to-noise ratios (SNR) are summarized in Table~\ref{tab:observation-summary}. The third night suffers from the lowest SNR among the five nights due to poor seeing conditions.
We exclude the first frame from the first night due to improper positioning on the slit. On the fifth night, 10 frames are discarded from further analysis as the star fell outside the slit owing to a telescope guiding error. \par

\begin{deluxetable*}{cccccccc}
\tablewidth{0pt}
\tablecaption{Summary of KELT-2A~b observation taken with IGRINS-2 \label{tab:observation-summary}}
\tablehead{
\colhead{Date} & \multirow{2}{*}{Airmass} & \multirow{2}{*}{Phase} & \colhead{Velocity relative to Earth} & \colhead{No. of Frames} & \colhead{Exp. Time}\tablenotemark{a} & \colhead{SNR}\\[-6pt]
in UTC & & & (km/s) & (Total/Used) & (s) &  at $1.66\, \mu \rm m$\tablenotemark{b} }
\startdata
{2025/02/21} & 1.03 -- 1.02 -- 1.79 & 0.421 -- 0.468 & $47.8$ -- $7.8$ & 61/60 & 91.45 & 255\\
{2025/03/10} & 1.02 -- 1.05 & 0.547 -- 0.558 & $-60.0$ -- $-69.8$ & 15/15 & 91.45 & 211\\
{2025/03/14} & 1.05 -- 1.96 & 0.527 -- 0.562 & $-42.2$ -- $-72.2$ & 44/44 & 91.45 & 128\\
{2026/01/12} & 1.49 -- 1.02 -- 1.08 & 0.421 -- 0.471 & $32.3$ -- $-10.5$ & 52/52 & 120 & 288\\
{2026/01/29} & 1.15 -- 1.02 -- 1.30 & 0.557 -- 0.606 & $-79.9$ -- $-118.7$ & 52/42 & 120 & 323\\
\enddata
\tablenotetext{a}{{Times listed are per exposure. Each frame is an AB-coadded pair, with effective integration time twice the listed value.}}
\tablenotetext{b}{{per pixel value}}
\end{deluxetable*}
\subsection{Data Reduction and Data Selection}
\label{subsec:data_selection}
All raw data were reduced to 1D extracted spectra for each order using the IGRINS Pipeline Package (PLP) Version 3.2, which was developed for IGRINS-2 \citep{IGRINS-2_PLP}. Due to the high SNR, some pixels were erroneously flagged as cosmic-ray hits by the standard pipeline and removed, resulting in slight spurious features in the spectra. Therefore, we did not apply cosmic-ray masking for the first, fourth, and fifth nights, which had high SNR. Pixels actually hit by cosmic rays are expected to have locally elevated counts; most such pixels are removed by the outlier mask described later.\par
We exclude orders 0, 1, and 24 – 26 of the \textit{H}-band ($1.836–1.793~$\micron, and $1.485–1.442~$\micron) and orders 0, 1, 19, and 21 - 25 of the \textit{K}-band ($2.555–2.486~$\micron, $2.017–1.990~$\micron, and $1.974–1.867~$\micron) from our analysis for all five nights due to low sensitivity and low telluric transmittance. 
Additionally, the extracted spectra from the fifth night exhibited unexplained jagged artifacts. Visual inspection revealed that the jagging was severe in orders 2 to 7 of the  \textit{H}-band ($1.800–1.694~$\micron), and these orders are excluded from further analysis. We notice that the edges of the blaze functions in several orders exhibit variations unrelated to airmass changes. These variations would be decomposed into more singular values (SVs) in the subsequent singular value decomposition (SVD) analysis, which consequently leads to the degradation of planetary signals. Therefore, following \cite{Line2021}, we exclude 100 pixels from each end of every order from our analysis. \par
Next, we identify outlier pixels and correct them. This is done as follows. First, for each exposure and each order, the spectrum is divided by the median spectrum taken along the time axis to normalize the continuum level. Each spectrum is then divided by its own median. In these normalized spectra, pixels deviating more than $5\sigma$ from the median value are flagged as outliers. After removing the flagged outliers, the same procedure is repeated to search for new outliers, and this process is iterated until no new outlier pixels are identified.
As in \cite{Brogi_2014} and \cite{Pelletier_2021}, a single isolated outlier pixel is corrected by spline interpolation along the wavelength axis. Two or three consecutive outlier pixels are corrected by linear interpolation along the wavelength axis, while a group of four or more consecutive outlier pixels is entirely masked with NaN. Finally, the spectra are multiplied by the median values along both the wavelength and time axes to restore the original continuum level. For subsequent processing, we create a common pixel mask by removing any pixel that is flagged as NaN in any of the frames from all frames.\par
\begin{figure}[ht!]
\plotone{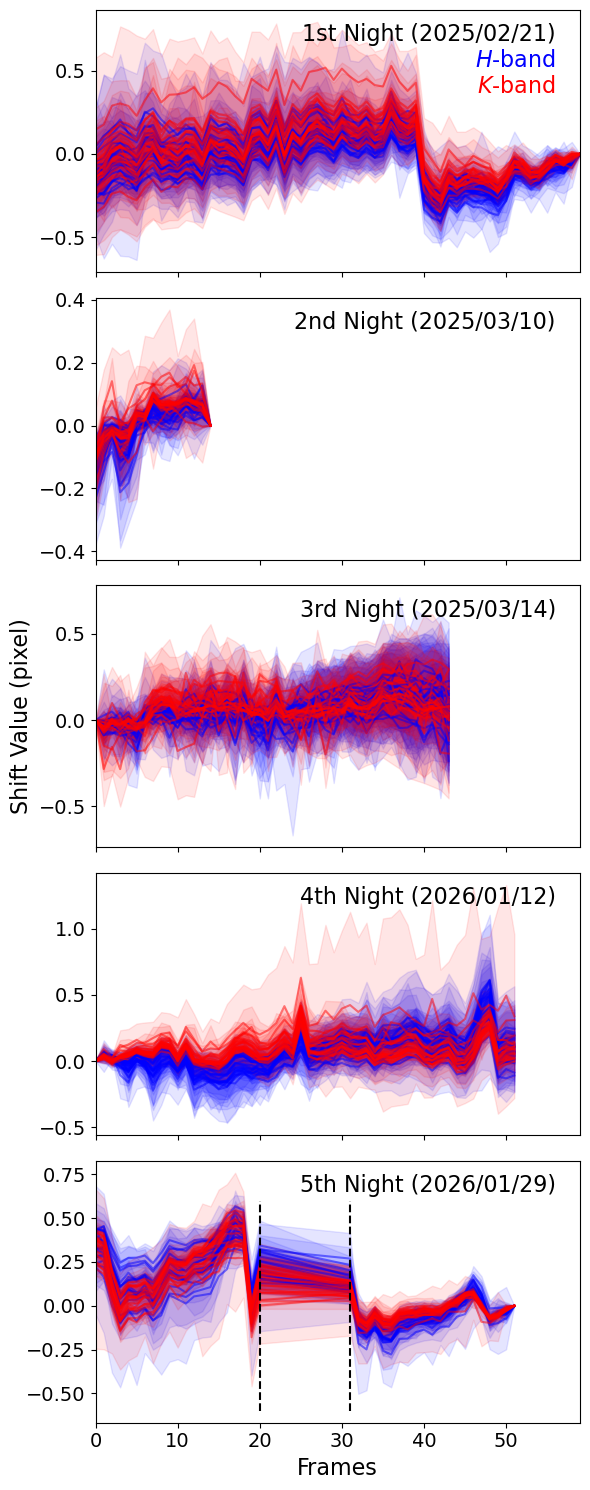}
\caption{{Wavelength shifts measured in pixels as a function of frame number for \textit{H}-band (blue) and \textit{K}-band (red) spectral orders across five observing nights. The shift is modeled as a linear function of wavelength, independently fit for each frame and order; solid lines show the mean shift for each order, and shaded regions show the standard deviation of the shift across the order, reflecting the wavelength dependence introduced by the slope of the linear fit. The dashed vertical lines in the fifth night (2026/01/29) mark the range of frames removed due to the target drifting off the slit.}}
\label{fig:shift_values}
\end{figure}

Similar to IGRINS, IGRINS-2 is a spectrograph without stabilized temperature, pressure, or slit illumination. As a result, the spectra drift $\sim 1$ pixel on the detector over time, necessitating additional wavelength alignment. Following previous studies \citep{Line2021, Brogi2023, Kanumalla2024, Mansfield2024, Smith2024_wasp77Ab, Smith2024_wasp121b, Wardenier2024, Panwar2025, Choi_2025, Sanchez_2026}, we perform this alignment as described below. 
IGRINS-2 conducts wavelength calibration each night using sky emission lines captured in zenith sky frames. {We align the spectra of all other frames to the wavelength solution of the frame taken closest in time to the sky frame acquisition, by shifting and stretching them -- i.e., applying a linear function of wavelength -- independently for each order and frame.} This reference frame corresponds to the last frame for the first, second, and fifth nights, and the first frame for the third and fourth nights. The shift values for each night are shown in Figure~\ref{fig:shift_values}.

Next, we remove pixels that vary significantly along the time axis. {This prevents a few problematic pixels from affecting the entire order and improves the subsequent telluric removal.} Specifically, for each order, the spectra are divided by the median along both the pixel and time axes, after which the standard deviation along the time axis is computed for each pixel. For each order, the median absolute deviation (MAD) of the standard deviations computed across all pixels is calculated, and pixels deviating by more than a factor of five are removed. The removed pixels include those located in deep telluric absorption regions and low SNR pixels near the edges of each order.\par

\subsection{Removing Telluric and Stellar Contamination}
\label{subsec:svd}

The observed spectra are dominated by stellar light and telluric absorption, obscuring planetary signals. Since these components are stationary in the wavelength direction, applying SVD along the time axis concentrates these components into the first few singular vectors with large singular values \citep{deKok2013}. 
The SVD approach decomposes the data matrix of the $i$-th order ${\bf A}_i$ -- with a size of $N_{\rm frame} \times N_{\rm pixel}$, where $N_{\rm frame}$ and $N_{\rm pixel}$ are the number of frames per night and the number of pixels in the $i$-th order, respectively -- into a product of three matrices:
\begin{equation}
    {\bf A}_i = {\bf U} {\bf S} {\bf V}^{\rm T},
\end{equation}

where $\bf U$ and $\bf V$ are orthogonal matrices with sizes of $N_{\rm frame} \times N_{\rm frame}$ and $N_{\rm pixel} \times N_{\rm pixel}$, respectively. $\bf S$ is a diagonal matrix with SVs arranged in descending order. We reconstruct a matrix ${\bf A}_i'$ by retaining only the first $k$ SVs:
\begin{equation}
\label{eq:low-order_approx}
    {\bf A}_i' = {\bf U}_k {\bf S}_k {\bf V}_k^{\rm T}, 
\end{equation}
where ${\bf U}_k$, ${\bf S}_k$, and ${\bf V}_k$ contain only the first $k$ components of $\bf U$, $\bf S$, and $\bf V$, respectively. By dividing the original data matrix ${\bf A}_i$ by the reconstructed spectra ${\bf A}_i'$, we obtain the normalized matrix $\tilde{{\bf A}}_i$ that isolates the planetary signals while removing the dominant stellar and telluric contributions:
\begin{equation}
    \tilde{{\bf A}}_i = {\bf A}_i \oslash {\bf A}_i',
\end{equation}
where $\oslash$ denotes element-wise division.\par
The number of SVs removed in the reconstruction is often chosen arbitrarily. {Here we adopt the minimum number that leaves no visually discernible telluric residuals, so as to avoid unnecessary attenuation of the planetary signal.
This criterion yields 3 SVs for Nights 1--4, while for Night 5 we remove 4 SVs, as clear telluric residuals persist with 3 SVs {(Figure~\ref{fig:trail})}.

\begin{figure*}[ht!]
\plotone{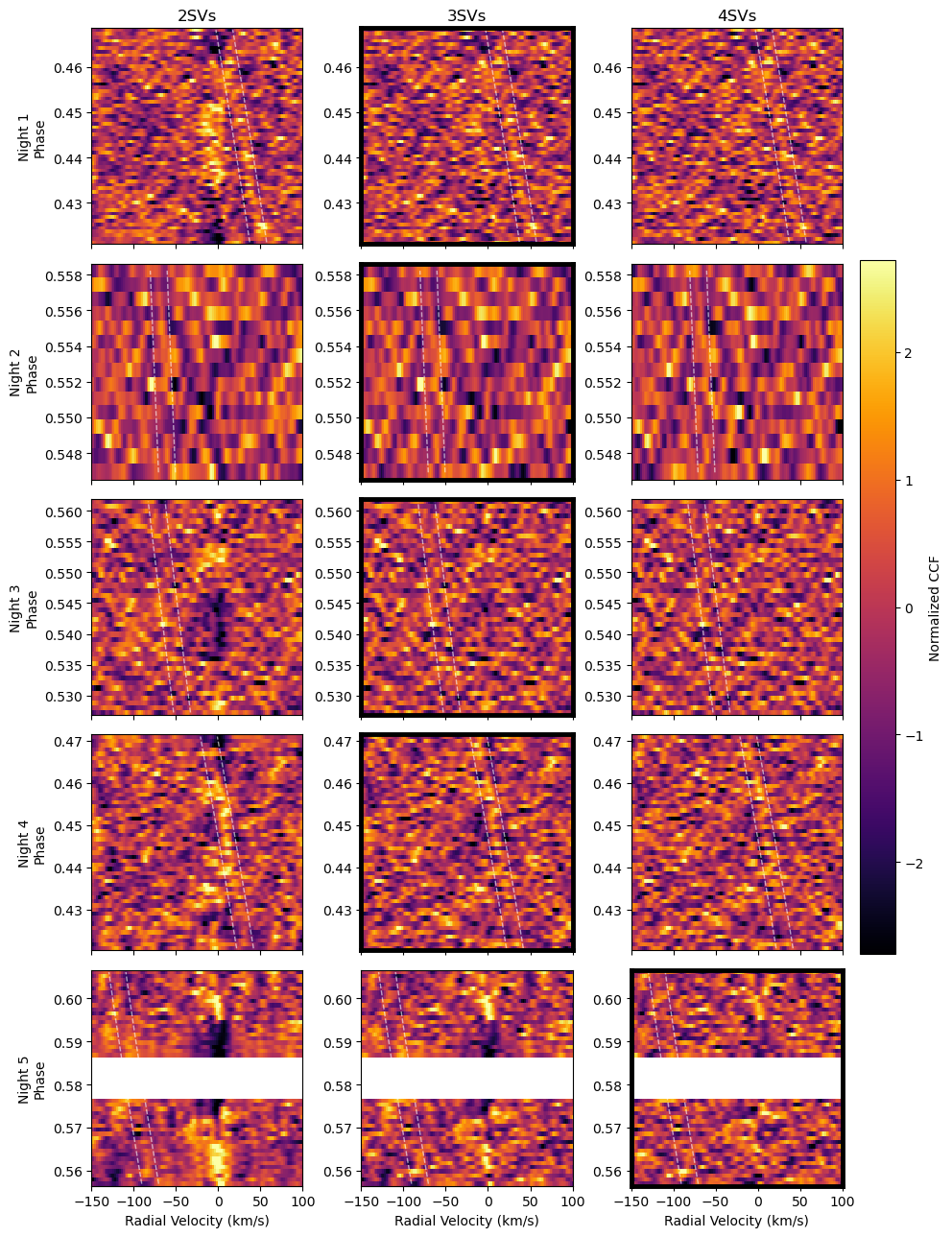}
\caption{Cross-correlation functions between the SVD-detrended spectra and the continuum-normalized best-fit planetary model, shown as a function of orbital phase and velocity in the Earth's rest frame, for 2, 3, and 4 SVs removed (left to right). Each row (frame) is normalized by subtracting its mean and dividing by its standard deviation. Residual telluric features appear as vertical structures near 0 km s$^{-1}$. White dashed lines bracket the expected planetary trail; the planetary signal is only faintly visible on Nights 1 and 4. Panels with thick frames indicate the number of SVs adopted in our analysis (3 for Nights 1–4 and 4 for Night 5). The blank region on Night 5 corresponds to the frames removed due to the guiding error (Section \ref{subsec:observation}).}
\label{fig:trail}
\end{figure*}

We confirm that the main results remain consistent within $\mathrm 1-2\sigma$ when the number of removed SVs is increased by up to three for each night (i.e., 3--6 SVs for Nights 1--4 and 4--7 SVs for Night 5), although the uncertainties on the retrieved parameters increase with the number of removed SVs. We discuss this behavior in detail in Section~\ref{sec:validation}.} \par

We store the divided spectrum and its wavelengths, and additionally, to accelerate the atmospheric retrieval described in Section \ref{subsec:reprocessing}, we store the reconstructed spectrum ${\bf A}_i'$ and ${\bf B}_i = {\bf U}_k({\bf U}_k^{\rm T}{\bf U}_k)^{-1}{\bf U}_k^{\rm T}$, as proposed by \cite{Gibson_2022}.\par
Subsequently, we mask outliers exceeding $3\, \sigma$ in each spectrum after dividing by the SVD-reconstructed spectrum. If a specific pixel is masked in more than 10\% of the frames for that night, we mask the corresponding pixel across all frames for that night.\par
Finally, for each exposure and each order, low-frequency continuum variations are removed by dividing by the spectrum convolved with a Gaussian kernel of FWHM = 80 pixels along the wavelength axis. 
We note that the results agree within $1\,\sigma$ even without this high-pass filtering procedure.

\section{Atmospheric Retrieval}
\label{sec:Atmospheric_Retireval}
To extract information on the temperature structure and molecular abundances from the data obtained in Section~\ref{section:Observation_Reduction}, we construct a spectral model that enables direct comparison with the data and perform atmospheric retrieval. 

\subsection{Atmospheric Modeling}
\label{subsection:atmospheric_modeling}
We use \texttt{petitRADTRANS} version 3 as the radiative transfer code to compute the planetary emission model spectra \citep{Molliere_2019, Blain_2024}. We assume a cloud-free atmosphere, as KELT-2A b has a sufficiently high equilibrium temperature that clouds are not expected to form on the dayside \citep{Parmentier_2016}, and the observed flux is expected to originate predominantly from cloud-free regions. The pressure range from $10^{-6}$ bar to $10^2$ bar is divided into 130 layers evenly spaced in log space. We begin with a free chemistry approach in which each atmospheric species has a constant vertical abundance, with H$_2$ and He serving as filler gases (He/H$_2$ number ratio = 0.176; \citealt{Line2021}) to normalize the total abundance to unity. We also perform an additional retrieval assuming chemical equilibrium, using \texttt{easyCHEM} \citep{Lei_Molliere_2025} with [C/H] and [O/H] as free parameters instead of sampling individual chemical species (hereafter EQ model). We adopt the temperature-pressure ($T$--$P$) profile parameterization of \cite{Guillot_2010}, and test alternative $T$--$P$ profile parameterizations in Section~\ref{subsec:retrieval_robustness}.\par

\subsubsection{Line-by-line Model for IGRINS-2}
\label{subsubsection:lbl_model}
Spectra over the wavelength range of $1.35$--$2.63\,\mu$m, which adequately covers the IGRINS-2 wavelength range, are computed at a resolution of $R=250,000$ in line-by-line mode. We consider $\rm H_2O$, $\rm ^{12}CO$, $\rm ^{13}CO$, and $\rm ^{12}CO_2$ as line opacity sources. The former three species have strong absorption bands in the IGRINS-2 wavelength range. Although $\rm ^{12}CO_2$ absorption bands are heavily contaminated by deep telluric $\rm CO_2$ absorption and are largely masked during data reduction as described in Section~\ref{subsec:data_selection}, we include it for consistency with the correlated-$k$ model used for the joint fit with \textit{Spitzer}/IRAC photometry written in Section~\ref{subsubsection:ck_model}.\par For opacity files, we use the POKAZATEL linelist \citep{Polyansky_2018} for $\rm H_2O$ and HITEMP database \citep{Rothman_2010} for $\rm ^{12}CO_2$, which are available as standard in \texttt{petitRADTRANS}. For $\rm ^{12}CO$ and $\rm ^{13}CO$, we download each partition functions and line lists from the HITEMP database \citep{Rothman_2010, Gordon_2022, Li_2015}, and calculate opacities using \texttt{HELIOS-K} \citep{Grimm_2015, Grimm_2021} with a resolution of $\rm 0.001\,cm^{-1}$, adopting Voigt profiles with a $\rm 100\,cm^{-1}$ cutoff imposed on the line wings, across the 200 temperature and pressure grids employed by \texttt{petitRADTRANS}. Since the spectral line intensities provided by HITEMP are pre-multiplied by natural terrestrial isotopic abundance ratios, we divide by the respective isotopic abundance ratios to normalize the line intensities to a per-molecule basis.
{We adopt the air-broadening parameters provided by HITEMP, although H$_2$-broadening parameters are now available for several molecules \citep{Tan_2022}}. \citet{Yama_2026} found that changing to broadening parameters under the $\rm H_2$ atmosphere introduces a $\sim0.05$ systematic offset in the derived C/O ratio for the brown dwarf binary Luhman 16AB. However, since the SNR of our dataset is lower than that of \cite{Yama_2026}, our C/O uncertainties exceed this systematic, and the impact on our conclusions is limited.
We also consider continuum opacity of H$_2$--H$_2$ collision induced absorption (CIA) from \cite{Borysow_2001, Borysow_2002} and H$_2$--He CIA from \cite{Borysow_1988, Borysow_1989a, Borysow_1989b}.\par
We convolve the calculated planetary model spectrum with a Gaussian kernel to match the instrumental wavelength resolution ($R=45,000$). Subsequently, we compute $v_p$, the velocity difference between Earth and the planet according to the following equation:
\begin{equation}
\label{eq:orbital_velocity}
v_p = K_{\rm p}\sin(2\pi\phi)+v_{\rm sys}-v_{\rm bary},
\end{equation}
where $K_{\rm p}$ is the Keplerian velocity, $\phi$ is the orbital phase, $v_{\rm sys}$ is the systemic velocity, and $v_{\rm bary}$ is the Earth's barycentric velocity calculated using \texttt{PyAstronomy.barycorr}. We calculate $\phi$ in Equation~(\ref{eq:orbital_velocity}) using the latest ephemeris from \cite{Kokori_2023} and Doppler-shift the planetary model wavelengths according to the derived $v_p$.\par
Finally, the model wavelengths are linearly interpolated to match the observed data wavelengths.

\subsubsection{Correlated-$k$ Model for \textit{Spitzer}/IRAC}
\label{subsubsection:ck_model}
For the joint fit with the \textit{Spitzer}/IRAC secondary-eclipse measurements in Channels 1 $\rm (3.6\, \mu m)$ and 2 $\rm (4.5\, \mu m)$ provided by \cite{Piskorz_2018}, we compute an additional planetary emission model using the correlated-$k$ method. Spectra are computed over the wavelength range of $3$--$5.5\,\mu$m at a resolution of $R=1,000$, which adequately covers the \textit{Spitzer}/IRAC Channel 1 and 2 bandpasses. We consider $\rm H_2O$, $\rm ^{12}CO$, $\rm ^{13}CO$, and $\rm ^{12}CO_2$ as line opacity sources. For $\rm H_2O$, $\rm ^{12}CO$, and $\rm ^{13}CO$, we use the same line lists as described in Section~\ref{subsubsection:lbl_model}. For $\rm ^{12}CO_2$, we use the linelist from the ExoMol database \citep{Yurchenko_2020} instead of HITEMP. We note that this inconsistency is unlikely to introduce significant differences in our results, given the limited spectral resolution and data quality of the \textit{Spitzer} photometric data points. The same continuum opacity sources as described in Section~\ref{subsubsection:lbl_model} are adopted.\par
From the computed planetary spectrum $F_{\rm p}$, we calculate the eclipse 
depth $D$ for each \textit{Spitzer}/IRAC channel using the subarray transmission 
curve $\mathcal{T}$ provided by \cite{Quijada_2004} as follows:
{
\begin{equation}
\label{eq:ecl_dep}
D = \frac{\int \mathcal{T} F_p(\lambda)\,d\lambda}
         {\int \mathcal{T} F_s(\lambda)\,d\lambda}
    \left(\frac{R_p}{R_s}\right)^2,
\end{equation}
where $F_s$ is the stellar surface flux taken from the PHOENIX model grid
of \cite{Husser_2013} at the stellar effective temperature $T_{\rm eff}$}, and
$R_p$ and $R_s$ are the planetary and stellar radii, respectively, for which we
adopt the values listed in Table~\ref{tab:planet_parameters}.

\subsection{Model Reprocessing}
\label{subsec:reprocessing}
\cite{Brogi_Line_2019, Gibson_2022} demonstrated that detrending processes can alter the planetary signal. In the case of SVD, this occurs because part of the planetary signal is distributed into the dominant vectors used for reconstruction, leading to signal alteration when the data is divided by the reconstructed data. Therefore, to simulate this signal alteration, we construct the model which includes stellar light and telluric contamination and reprocess this model by applying the same operations as those applied to the data.\par

We assume that the observed spectrum $F_{\rm obs}(\lambda, t)$ can be described as follows:
\begin{equation}
\label{eq:F_obs}
F_{\rm obs}(\lambda, t) = (1 + F_{\rm scaled}(\lambda, t))F_s(\lambda) T(\lambda, t),
\end{equation}
where $T(\lambda, t)$ is the throughput from Earth's atmosphere to the detectors. $F_{\rm scaled}(\lambda, t)$ is the planetary spectrum scaled by the stellar spectrum, described as:
\begin{equation}
\label{eq:F_scaled}
F_{\rm scaled}(\lambda, t) = \frac{F_p(\lambda, t)}{F_s(\lambda)}\left( \frac{R_p}{R_s}\right)^2,
\end{equation}
where $F_p(\lambda, t)$ and $F_s(\lambda)$ are the model spectra of planet and star, respectively. $R_p$ and $R_s$ are the planetary and stellar radii, respectively. \par
For model construction, we define each variable in Equations (\ref{eq:F_obs})
and (\ref{eq:F_scaled}) as follows: $F_p(\lambda, t)$ is the planetary spectrum
calculated in Section \ref{subsection:atmospheric_modeling}; {$F_s(\lambda)$ is
the stellar surface flux taken from the PHOENIX model grid of
\cite{Husser_2013} at the stellar effective temperature $T_{\rm eff}$; and
$F_s(\lambda) T(\lambda, t)$ corresponds to the reconstructed spectra
${\bf A}'$ saved in Section \ref{subsec:svd}. We adopt the values of $R_p$,
$R_s$, and $T_{\mathrm{eff}}$ from Table~\ref{tab:planet_parameters}.}
Therefore, the model matrix $\bf{M}$ is given by:
\begin{equation}
    {\bf M} = \left\{1+\frac{F_p(\lambda, t)}{F_s(\lambda)}
    \left(\frac{R_p}{R_s}\right)^2\right\}\odot{\bf A}',
\end{equation}
where $\odot$ denotes element-wise multiplication.\par
Instead of decomposing $\bf M$ through SVD and reconstructing the data matrix $\bf{M}'$, we approximate this process by multiplying the saved matrix ${\bf B}$ from Section \ref{subsec:svd} on the left, following the approach of \cite{Gibson_2022} and \cite{Dash_2024}:
\begin{equation}
{\bf M}' = {\bf B} {\bf M}.    
\end{equation}
This method reduces the computational time required for model reprocessing. Subsequently, we divide $\bf M$ by the reconstructed one, 
similar to the data processing, to obtain a reprocessed model $\tilde{{\bf M}}$ 
that can be directly compared with the data matrix $\tilde{\bf A}$:
\begin{equation}
    \tilde{{\bf M}} = {\bf M} \oslash {\bf M}'.
\end{equation}

\subsection{Parameter Estimation}
\label{subsec:parameter_estimate}
The free chemistry and EQ models have 11 and 10 sampled parameters, respectively, as listed in Table~\ref{tab:parameter_prior_posterior}. While \texttt{petitRADTRANS} uses mass-mixing ratios (MMR) for chemical abundances, we sample volume-mixing ratios (VMR) using the following relation:
\begin{equation}
    X_i = \frac{\mu_i}{\mu} n_i,
\end{equation}
where $X_i$, $n_i$, and $\mu_i$ are the MMR, VMR, and molar mass of species $i$, respectively. $\mu$ is the mean molecular weight 
of the atmosphere.\par

\begin{deluxetable}{lccccccc}
\tablewidth{0pt}
\tablecaption{List of Parameters, Priors, and Posteriors for Atmospheric Retrieval \label{tab:parameter_prior_posterior}}
\tablehead{
\multirow{2}{*}{Parameter Name} & \multirow{2}{*}{Symbol} & \multicolumn{2}{c}{Prior} & \multicolumn{2}{c}{Posterior (IGRINS-2 only)} & \multicolumn{2}{c}{Posterior (IGRINS-2 + IRAC)}\\
\cline{3-8}
& & \colhead{Free} & \colhead{EQ} & \colhead{Free} & \colhead{EQ} & \colhead{Free} & \colhead{EQ}}
\startdata
log of H$_2$O gas VMR &  log(H$_2$O) & $\mathcal{U}{(-12, -0.1)}$ & - & $-3.57^{+0.29}_{-0.21}$ & - & $-3.78^{+0.27}_{-0.20}$ & - \\
log of $^{12}$CO gas VMR & log($^{12}$CO) & $\mathcal{U}{(-12, -0.1)}$ & - & $-3.98^{+0.41}_{-0.36}$ & - & $-4.17^{+0.39}_{-0.32}$ & -\\
C to H ratio (relative to solar, dex) & [C/H] & -  & $\mathcal{U}(-3, 3)$ & - & $-0.58^{+0.44}_{-0.36}$ & - & $-0.81^{+0.41}_{-0.34}$ \\
O to H ratio (relative to solar, dex) & [O/H] & -  & $\mathcal{U}(-3, 3)$ & - & $-0.12^{+0.28}_{-0.23}$ & - & $-0.37^{+0.27}_{-0.17}$ \\
log of $^{12}$CO/$^{13}$CO & log($^{12}$CO/$^{13}$CO) & \multicolumn{2}{c}{$\mathcal{U}(-4, 4)$} & $>1.48 \, (97.5\%)$ & $>1.45 \, (97.5\%)$ & $>1.37 \, (97.5\%)$ & $>1.33 \, (97.5\%)$ \\
log of $^{12}$CO$_2$ gas VMR &  log($^{12}$CO$_2$) & $\mathcal{U}(-12, -0.1)$ & - & $<-5.13 \, (97.5\%)$ & - & $<-5.43 \, (97.5\%)$ & -\\
log of infrared opacity  ($\rm cm^2\ g^{-1}$) & $\log(\kappa_{\rm IR})$ & \multicolumn{2}{c}{$\mathcal{U}(-3, 3)$} & $-0.85^{+0.54}_{-0.34}$ & $-0.74^{+0.52}_{-0.34}$ & $-1.17^{+0.34}_{-0.24}$ & $-1.01^{+0.38}_{-0.26}$\\
log of optical-infrared opacity ratio & $\log(\gamma)$ & \multicolumn{2}{c}{$\mathcal{U}(-3, 3)$} & $-1.00^{+0.23}_{-0.44}$ & $-1.00^{+0.24}_{-0.39}$ & $-0.74^{+0.12}_{-0.18}$ & $-0.79^{+0.15}_{-0.24}$\\
Intrinsic temperature (K) & $T_{\rm int}$ & \multicolumn{2}{c}{$\mathcal{U}(10,  500)$} & $<380 \, (97.5\%) $ & $<375 \, (97.5\%) $ & $<450 \, (97.5\%) $ & $<420 \, (97.5\%) $\\
Equilibrium temperature (K) & $T_{\rm eq}$ & \multicolumn{2}{c}{$\mathcal{U}(500, 2000)$} & $1190^{+170}_{-260}$ & $1190^{+180}_{-240}$ & $1410^{+100}_{-130}$ & $1380^{+110}_{-160}$ \\
$K_{\rm p}$ offset\tablenotemark{a} (km/s) & $dK_{\rm p}$ & \multicolumn{2}{c}{$\mathcal{U}(-20, 20)$} & $0.00^{+0.57}_{-0.56}$ & $-0.03\pm0.56$ & $0.03^{+0.60}_{-0.56}$ & $0.08^{+0.58}_{-0.57}$\\
$v_{\rm sys}$ offset\tablenotemark{b} (km/s) & $dv_{\rm sys}$ & \multicolumn{2}{c}{$\mathcal{U}(-20, 20)$} & $0.42\pm0.22$ & $0.41^{+0.22}_{-0.20}$ & $0.45\pm0.22$ & $0.46\pm0.22$ \\
noise scaling factor & $\beta$ & \multicolumn{2}{c}{$\mathcal{U}(0.1, 2)$} & $0.6436\pm0.0001$ & $0.6436\pm0.0001$ & $0.6436\pm0.0001$ & $0.6436\pm0.0001$ \\
\enddata
\tablenotetext{a}{Measured relative to $K_p = 145~\rm{km~s^{-1}}$.}
\tablenotetext{b}{Measured relative to $v_{\rm sys}=-47.2\rm{km~s^{-1}}$ \citep{Gaia_2018}.}
\end{deluxetable}

To obtain the posterior distribution of these model parameters, we use the 
log-likelihood function introduced by \cite{Gibson_2020}:
\begin{equation}
\label{eq:log_likelihood_gibsonLL}
    {\ln (\mathcal{L_{\rm IGRINS-2}}) = -\frac{1}{2}\sum_i \frac{(a_i-m_i)^2}{(\beta \sigma_i)^2}-N\ln\beta},
\end{equation}
where $N$ is the number of data points excluding NaNs, $a_i$ and $m_i$ are the data points and reprocessed model, $\sigma_i$ is the error of each data point, and $\beta$ is an error scaling factor common to all nights. {We also tested a model in which $\beta$ was allowed to vary independently for each night. The resulting posteriors for all parameters other than $\beta$  agreed within $1\, \sigma$ with those from the common-$\beta$ model; we therefore adopt a single $\beta$ common to all nights for computational efficiency.}\par
Rather than computing the variance $\sigma_i^2$ directly from the data points as in \cite{Gibson_2020}, we instead use the variance provided by the IGRINS-2 PLP, divided by the same reconstructed spectra ${\bf A}'$ as applied to the data. \par
Note that the log-likelihood function in Equation~(\ref{eq:log_likelihood_gibsonLL}) does not account for correlated noise introduced by SVD. 
However, we confirm that this correlated noise is sufficiently small compared to the white noise at the current data quality, and does not significantly affect the interpretation of our results.
To verify this, we generated mock data with a SNR similar to the real observations outlined in Section~\ref{subsec:observation} and injected the best-fit model obtained in Section~\ref{subsec:retrieval_results}. We then repeated the retrieval hundreds of times with randomized noise realizations, with the same retrieval treatment as the real data, except that the temperature structure parameters were fixed to the injected values to speed up the sampling. We confirmed that the squared Mahalanobis distance between the retrieved parameters and the injected values follows a chi-squared distribution with degrees of freedom equal to the number of model parameters. However, a more rigorous verification is beyond the scope of this paper; {we plan to pursue the development of an HRCCS retrieval framework incorporating correlated noise in future work.}\par
The log-likelihood is computed according to Equation (\ref{eq:log_likelihood_gibsonLL}) 
for each night, each frame, and each order, and then summed. For the 
\textit{Spitzer}/IRAC photometry, we add the following log-likelihood:
\begin{equation}
\label{eq:log_likelihood_irac}
    \ln(\mathcal{L}_{\rm IRAC}) = -\frac{1}{2}\frac{(\delta_{\rm Ch1}-D_{\rm Ch1})^2}{\sigma_{\rm Ch1}^2}-\frac{1}{2}\frac{(\delta_{\rm Ch2}-D_{\rm Ch2})^2}{\sigma_{\rm Ch2}^2},
\end{equation}
where $\delta_{\rm Ch1}$ and $\delta_{\rm Ch2}$ are the observed eclipse depths in \cite{Piskorz_2018}, 
$D_{\rm Ch1}$ and $D_{\rm Ch2}$ are the model eclipse depths computed according 
to Equation~(\ref{eq:ecl_dep}), and $\sigma_{\rm Ch1}$ and $\sigma_{\rm Ch2}$ 
are the observational uncertainties for IRAC Channel 1 and Channel 2, respectively. 
The total log-likelihood is then given by the sum of all contributions:
\begin{equation}
    \ln(\mathcal{L}_{\rm total}) = \sum_{\rm nights}\ln(\mathcal{L_{\rm IGRINS-2}}) + \ln(\mathcal{L}_{\rm IRAC}).
\end{equation}\par
We use \texttt{PyMultiNest} \citep{Buchner_2014}, a Python wrapper for \texttt{MultiNest} \citep{Feroz_2007, Feroz_2009, Feroz_2019}, as our sampler. The prior distributions for each parameter are listed in Table~\ref{tab:parameter_prior_posterior}. We employ 500 live points and a convergence criterion of $\Delta\ln(\mathcal{Z}) < 0.1$.

\section{Results} \label{sec:results}

\subsection{$K_{\rm p}$ -- $v_{\rm sys}$ map}
\label{subsec:KpVsys_results}
\begin{figure*}[ht!]
\plotone{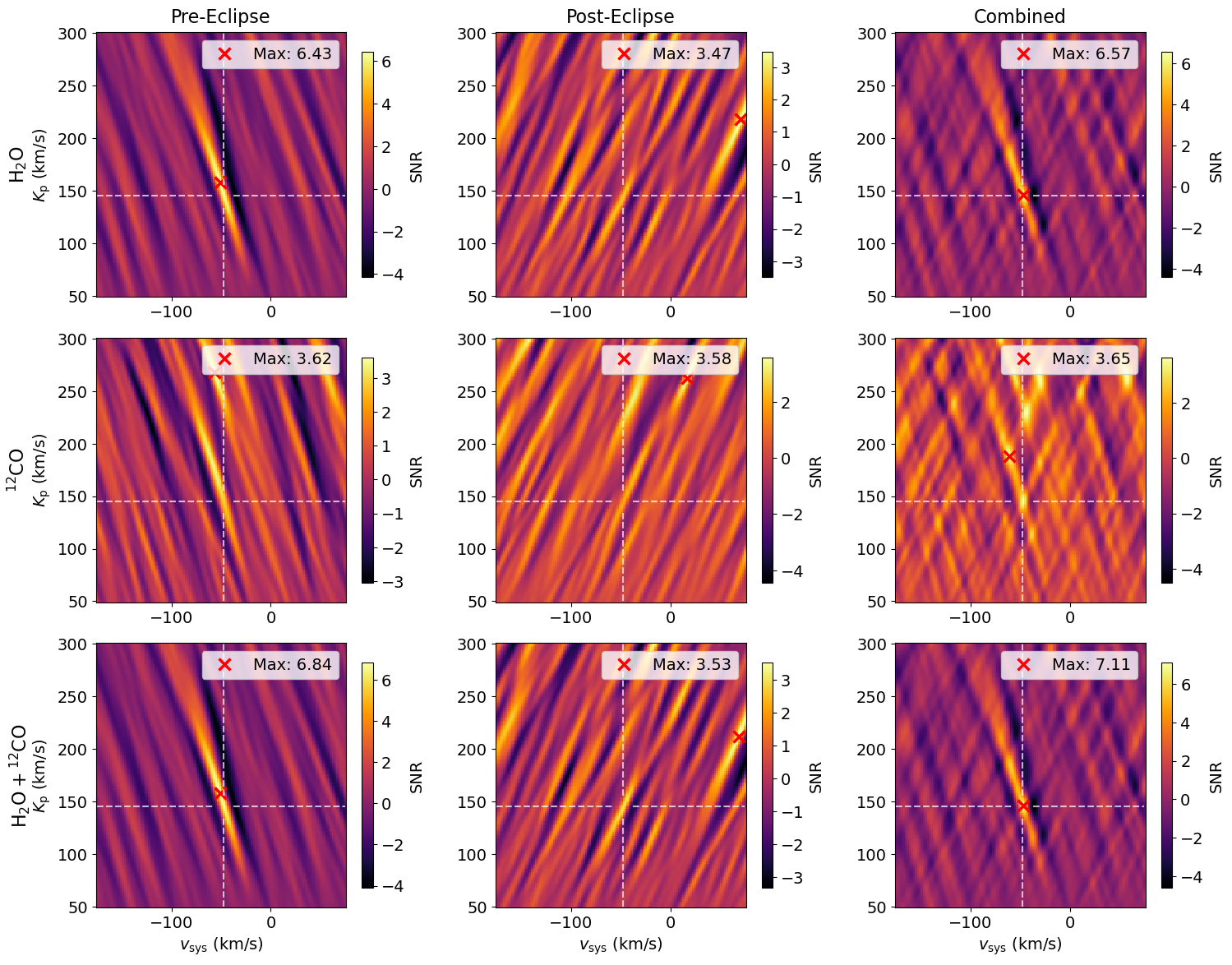}
\caption{CCF/SNR $K_{\rm p}$--$v_{\rm sys}$ maps. \textit{Left:} Calculated with pre-eclipse phase datasets. \textit{Mid:} Calculated with post-eclipse phase datasets.
\textit{Right:} Calculated with all datasets. Rows show 
H$_2$O (top), $^{12}$CO (middle), and H$_2$O $+$ $^{12}$CO model (bottom). White dashed 
lines indicate expected velocities; red crosses mark maximum values. Maximum SNR values are written in the upper right corner in each panel.}
\label{fig:SNR_KpVsys_map}
\end{figure*}

To visually assess the planetary signal, we compute the cross-correlation 
function (CCF) and log-likelihood by varying $K_{\rm p}$ and $v_{\rm sys}$ in 2~km~s$^{-1}$ steps, using the retrieved best-fit parameter values summarized in Table~\ref{tab:parameter_prior_posterior}. {For each combination of $K_{\rm p}$ and $v_{\rm sys}$, the planetary atmospheric model is Doppler-shifted accordingly and subjected to reprocessing (Section~\ref{subsec:reprocessing}).} The best-fit model is characterized by a non-inverted temperature structure and gas abundances {depleted} relative to solar composition, as detailed in Section~\ref{subsec:retrieval_results}. The CCF is calculated as:
\begin{equation}
    {\rm CCF} = \sum_i\frac{a_im_i}{\sigma_i^2}.
\end{equation}
Figure~\ref{fig:SNR_KpVsys_map} presents the results as $K_{\rm p}$ -- $v_{\rm sys}$ 
maps, where the CCF is converted to a detection 
signal-to-noise ratio (SNR) via: 
\begin{equation}
    {\rm SNR} = \frac{{\rm CCF} - {\rm med}}{{\rm std}},
\end{equation}
where $\rm med$ and $\rm std$ are the $3\, \sigma$-clipped median and standard deviation of the CCF values, respectively \citep{Kasper_2021, Mansfield2024}. While H$_2$O gas has absorption lines widely distributed across the observed wavelength range, CO gas has strong absorption lines only at $>2.3\,\mu\mathrm{m}$. The CO $K_{\rm p}$--$v_{\rm sys}$ map is computed using all orders, as limiting the analysis to only the six reddest orders covering wavelengths longer than $2.3\,\mu\mathrm{m}$ does not yield a significant improvement in SNR. \par
In the pre-eclipse phase, we find an H$_2$O signal at $\rm SNR = {6.43}$ near the expected velocity space, confirming the hints of H$_2$O absorption reported by \citet{Piskorz_2018} at a higher significance. For $^{12}$CO, we do not find a clear signal. A combined model including both H$_2$O and $^{12}$CO finds the planetary signal at SNR $= {6.84}$.\par
In the post-eclipse phase, no clear signal is found for either species. This is likely due to the limited data quality of the post-eclipse phase (see Section~\ref{subsec:Injection-Recovery}).\par
When combining all nights, H$_2$O is found at SNR $= {6.57}$, and a combined model including both H$_2$O and $^{12}$CO finds the planetary signal at SNR $= {7.11}$.
{Although $^{12}$CO is not significantly detected on its own, adding $^{12}$CO to the H$_2$O model consistently increases the SNR, both in the pre-eclipse phase and in the combined dataset. This indicates that the $^{12}$CO lines contribute coherently to the planetary signal, providing some evidence for the presence of $^{12}$CO in our data.}

\begin{figure*}[ht!]
\plotone{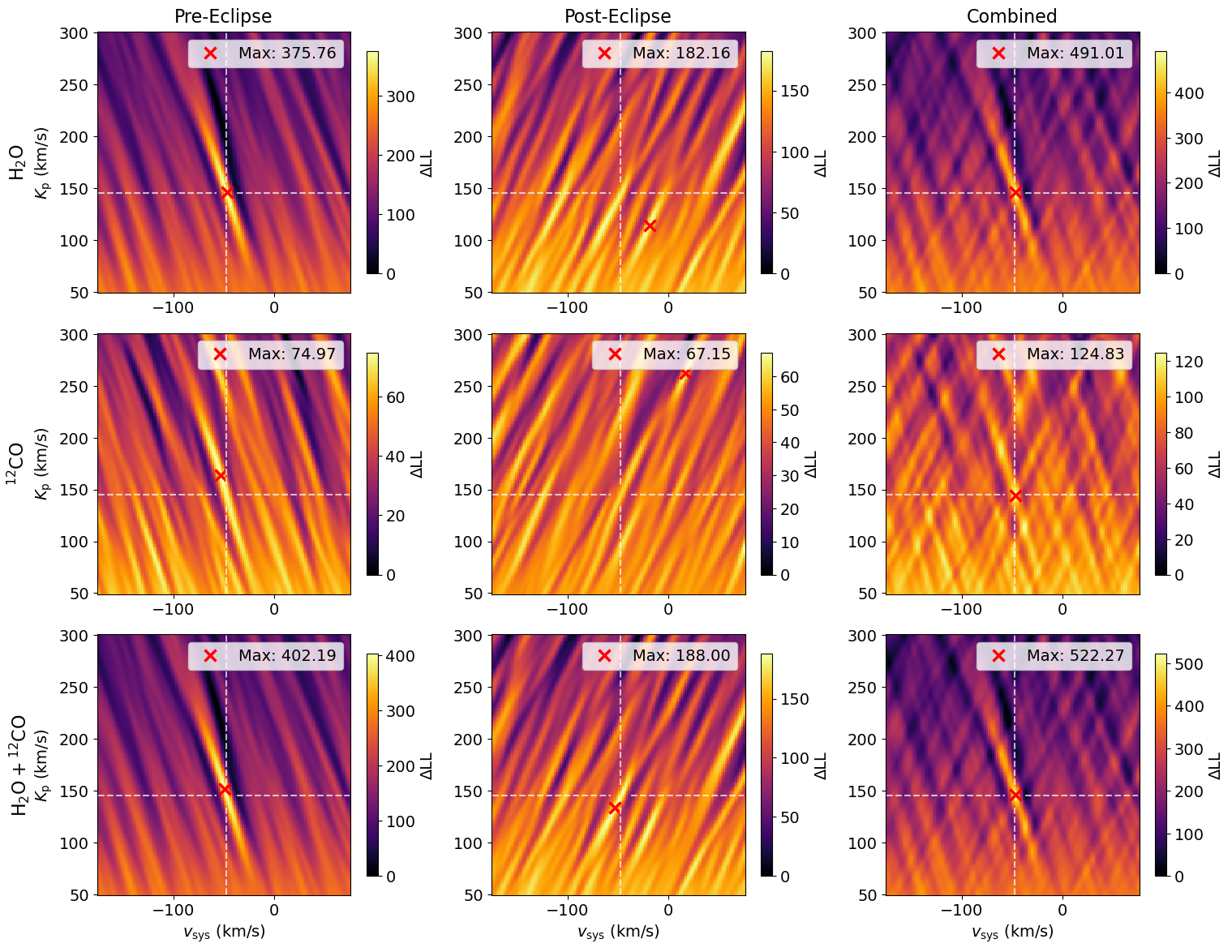}
\caption{Same as Figure~\ref{fig:SNR_KpVsys_map}, but map for $\Delta \ln (\mathcal{L})$.}
\label{fig:deltaLL_KpVsys_map}
\end{figure*}

Figure~\ref{fig:deltaLL_KpVsys_map} shows the $K_{\rm p}$ -- $v_{\rm sys}$ map of
\begin{equation}
    \Delta\ln(\mathcal{L}) = \ln(\mathcal{L_{\rm IGRINS-2}}) - \min(\ln(\mathcal{L_{\rm IGRINS-2}})),
\end{equation}
where $\ln(\mathcal{L_{\rm IGRINS-2}})$ is the log-likelihood computed according to 
Equation~(\ref{eq:log_likelihood_gibsonLL}), and $\min(\ln(\mathcal{L_{\rm IGRINS-2}}))$ is the minimum value of $\ln(\mathcal{L_{\rm IGRINS-2}})$ over all computed $K_{\rm p}$ and $v_{\rm sys}$ pairs.
The overall trend is similar to Figure~\ref{fig:SNR_KpVsys_map}, with H$_2$O signals detected in the pre-eclipse and combined datasets. In the $\Delta\ln(\mathcal{L})$ map, however, the maximum for $^{12}$CO is located near the expected planetary velocity space for {the pre-eclipse and} the combined datasets. This is because the log-likelihood is sensitive not only to the positions of absorption lines but also to their shapes (e.g., \citealt{Line2021, Brogi2023}). However, we do not claim a detection of $\rm ^{12}CO$. Future observations will be required to confirm the tentative $\rm ^{12}CO$ signal obtained here.

\subsection{Results of Retrieval Analysis}
\label{subsec:retrieval_results}
\begin{figure*}[ht!]
\plotone{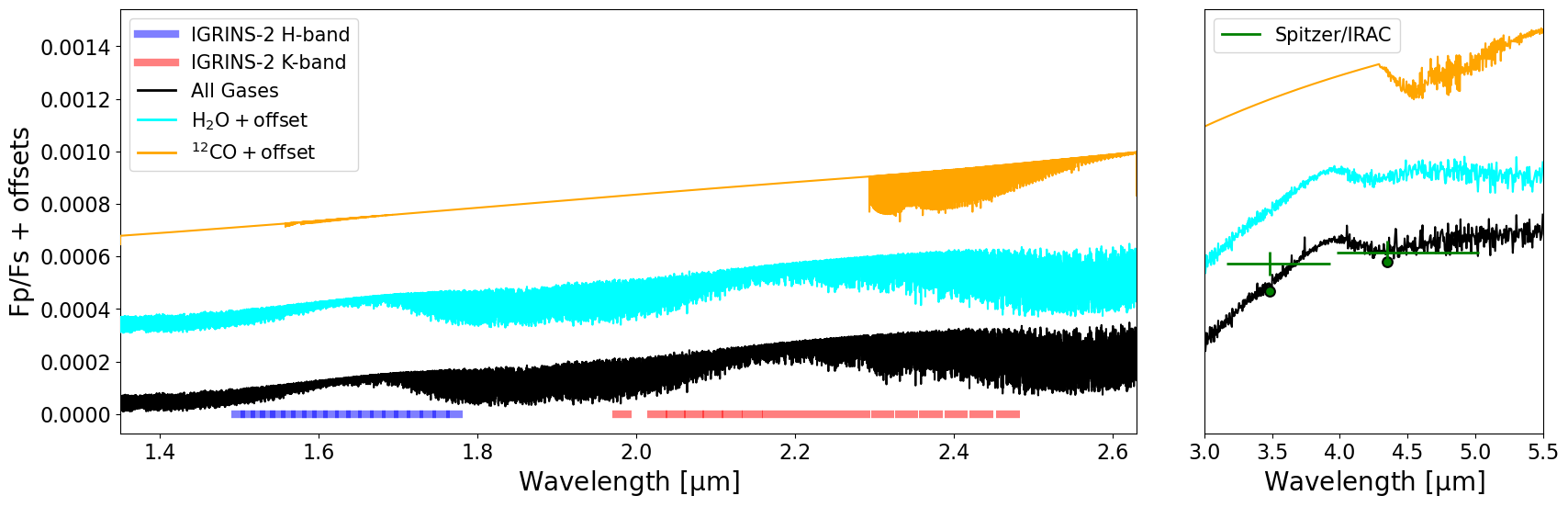}
\caption{\textit{Left}: Best-fit emission spectrum (black) from the EQ chemistry retrieval with the \citet{Guillot_2010} $T$--$P$ profile, including the \textit{Spitzer}/IRAC photometry. Cyan and orange lines show the spectra computed including only $\rm H_2O$ and $\rm ^{12}CO$, respectively, {each vertically offset for clarity.} Blue and red horizontal bars indicate the wavelength coverage of the IGRINS-2 \textit{H}- and \textit{K}-bands, respectively. \textit{Right}: Same as the left panel, but for the \textit{Spitzer}/IRAC wavelength range. Green data points with error bars show the \textit{Spitzer}/IRAC photometric measurements, while the filled green circles indicate the best-fit model spectrum integrated over the \textit{Spitzer}/IRAC bandpasses.}
\label{fig:best-fit_spectrum}
\end{figure*}

The posteriors calculated from the retrieval analysis in Section~\ref{sec:Atmospheric_Retireval} are summarized in Table~\ref{tab:parameter_prior_posterior}, the best-fit spectrum is shown in Figure~\ref{fig:best-fit_spectrum}, and Figure~\ref{fig:retrieved_summary} illustrates the retrieved $T$--$P$ profile, the emission contribution function, and the chemical abundance posteriors.
\subsubsection{$T$--$P$ profile}
\label{subsec:T-P_profile}


\begin{figure*}[ht!]
\plotone{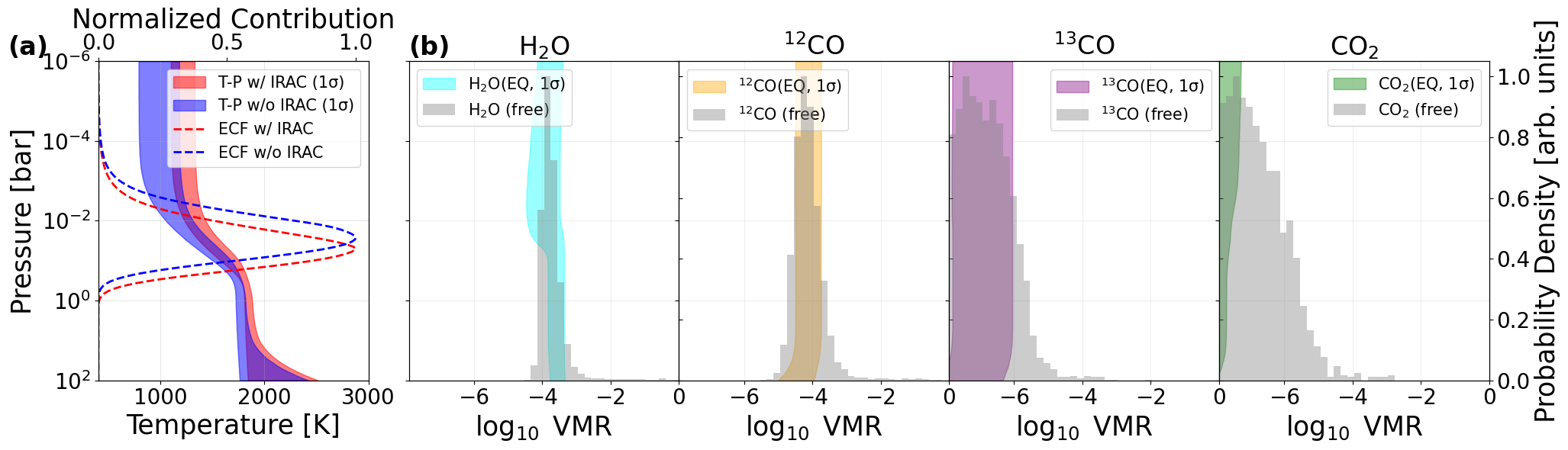}
\caption{The summary of retrieval results. (a): Retrieved temperature-pressure ($T$--$P$) profiles and emission contribution functions (ECFs) integrated over the IGRINS-2 wavelength range. Red and blue shaded regions show the 1$\sigma$ credible intervals of the $T$--$P$ profiles with and without the \textit{Spitzer}/IRAC photometry, respectively. Red and blue dashed curves show the corresponding ECFs. (b): Posterior distributions of the volume mixing ratios (VMRs) for H$_2$O, $^{12}$CO, $^{13}$CO, and $^{12}$CO$_2$ (from left to right), derived from the joint fit with the \textit{Spitzer}/IRAC photometry. Colored regions and gray histograms show the  1$\sigma$ credible intervals from the equilibrium chemistry retrieval and the posterior distributions from the free chemistry retrieval, respectively.}

\label{fig:retrieved_summary}
\end{figure*}

The retrieved $T$--$P$ profile {(Figure~\ref{fig:retrieved_summary}a)} shows a non-inverted structure around {$\sim$0.1~bar}, in agreement with the previous findings of \citet{Piskorz_2018} based on Keck/NIRSPEC and \textit{Spitzer} observations. This non-inverted structure is similar to that of other hot Jupiters with comparable equilibrium temperatures ($T_{\rm eq} \sim 1700$~K), including 
WASP-77A~b \citep{Line2021, August_2023, Smith2024_wasp77Ab, Edwards_Changeat_2024}, tau~Boo~b \citep{Brogi_2012, Pelletier_2021, Webb_2022, Panwar_2024}, and HD~149026~b \citep{Bean_2023}. \par
The inclusion of the two \textit{Spitzer}/IRAC photometric data points improves the constraints on the $T$--$P$ profile. This is because high-resolution spectroscopy is sensitive only to the contrast between continuum and lines, or between different lines, but not to the absolute continuum level, leading to a degeneracy between the temperature structure and the molecular mixing ratios. 
Combining with space-based photometry, which is sensitive to the absolute flux level, partially breaks this degeneracy \citep[e.g.,][]{Brogi_2017, Gandhi_2019, Yan_2022, Boucher_2023, Smith2024_wasp77Ab, Cont_2024}.\par
This constraint, however, is contingent on the absolute eclipse depth of the \textit{Spitzer} data being reliable, with no vertical offset such as those sometimes seen in \textit{HST}/WFC3 observations. For reference, the \textit{Spitzer} photometry for WASP-77A b appears to be in good agreement with the JWST emission spectroscopy presented in \citet{August_2023}, providing some reassurance regarding its reliability.

\subsubsection{Chemical Abundances}
\label{subsec:chemical_abundances}
{From the free chemistry retrieval,} we obtain bounded constraints on $\log(\rm H_2O)$ of ${-3.57^{+0.29}_{-0.21}}$ from IGRINS-2 alone, and ${-3.78^{+0.27}_{-0.20}}$ from the joint fit with \textit{Spitzer}/IRAC. For $\log(^{12}\rm CO)$, {we get $\log(\rm ^{12}CO)$ of ${-3.98^{+0.41}_{-0.36}}$ from IGRINS-2 alone, and ${-4.17^{+0.39}_{-0.32}}$ from the joint fit with \textit{Spitzer}/IRAC.} The IGRINS-2-only and joint fit results are consistent within $1\sigma$, and the inclusion of the \textit{Spitzer}/IRAC photometry marginally improves the constraints, as described in Section~\ref{subsec:T-P_profile}, by constraining the absolute continuum level and partially breaking the degeneracy between the temperature structure and the molecular abundances. The $^{12}$CO/$^{13}$CO ratio provides only a lower limit and $\log(^{12}\rm CO_2)$ provides only an upper limit. We confirmed that the other parameters remain consistent within $1\, \sigma$ even after excluding these two molecules ($^{13}\rm CO$ and $^{12}\rm CO_2$).\par
As discussed later, atmospheric elemental abundance ratios such as C/O and metallicity serve as proxies for the planet formation location (e.g., \citealt{Oberg_2011}). Therefore, we calculate the planetary C/O ratio 
and the [(C+O)/H] ratio as a proxy for [M/H] from the retrieved chemical 
abundances as follows:
\begin{equation}
\label{eq:C/O_calc}
{\rm C/O} =  \frac{n_{^{12}{\rm CO}} + n_{^{13}{\rm CO}}+n_{^{12}{\rm CO_2}}}{n_{^{12}{\rm CO}} + n_{^{13}{\rm CO}} +2n_{^{12}{\rm CO_2}} + n_{\rm H_2O}},
\end{equation}
and
\begin{equation}
\label{eq:[(C+O)/H]_calc}
\begin{split}
{\rm [(C+O)/H]} = {}& \log_{10} \left\{ \frac{2n_{^{12}{\rm CO}} + 2n_{^{13}{\rm CO}} + 3n_{^{12}{\rm CO_2}}+ n_{\rm H_2O}}{2n_{\rm H_2}+2n_{\rm H_2O}} \right\} \\
& - \log_{10} \left( \frac{{n_{\rm C}}_{\odot}}{{n_{\rm H}}_{\odot}}+\frac{{n_{\rm O}}_{\odot}}{{n_{\rm H}}_{\odot}} \right),
\end{split}
\end{equation}
where $n_i$ ($i = {}^{12}{\rm CO}$, ${}^{13}{\rm CO}$, ${}^{12}{\rm CO_2}$, H$_2$O, H$_2$) denotes 
the volume mixing ratio of each species, and ${n_{\rm C}}_\odot/{n_{\rm H}}_\odot$ 
and ${n_{\rm O}}_\odot/{n_{\rm H}}_\odot$ represent the solar carbon and oxygen 
abundances, respectively, for which we adopt the values from \citet{Asplund_2009}.
Similarly, we also calculate [C/H] and [O/H]:
\begin{equation}
\label{eq:[C/H]_calc}
\begin{split}
{\rm [C/H]} = {}& \log_{10} \left\{ \frac{n_{^{12}{\rm CO}} + n_{^{13}{\rm CO}}+n_{^{12}{\rm CO_2}}}{2n_{\rm H_2}+2n_{\rm H_2O}} \right\} \\
& - \log_{10} \left( \frac{{n_{\rm C}}_{\odot}}{{n_{\rm H}}_{\odot}} \right),
\end{split}
\end{equation}
and
\begin{equation}
\label{eq:[O/H]_calc}
\begin{split}
{\rm [O/H]} = {}& \log_{10} \left\{ \frac{n_{^{12}{\rm CO}} + n_{^{13}{\rm CO}} + 2n_{^{12}{\rm CO_2}} + n_{\rm H_2O}}{2n_{\rm H_2}+2n_{\rm H_2O}} \right\} \\
& - \log_{10} \left(\frac{{n_{\rm O}}_{\odot}}{{n_{\rm H}}_{\odot}} \right).
\end{split}
\end{equation}\par
\begin{figure}[ht!]
\plotone{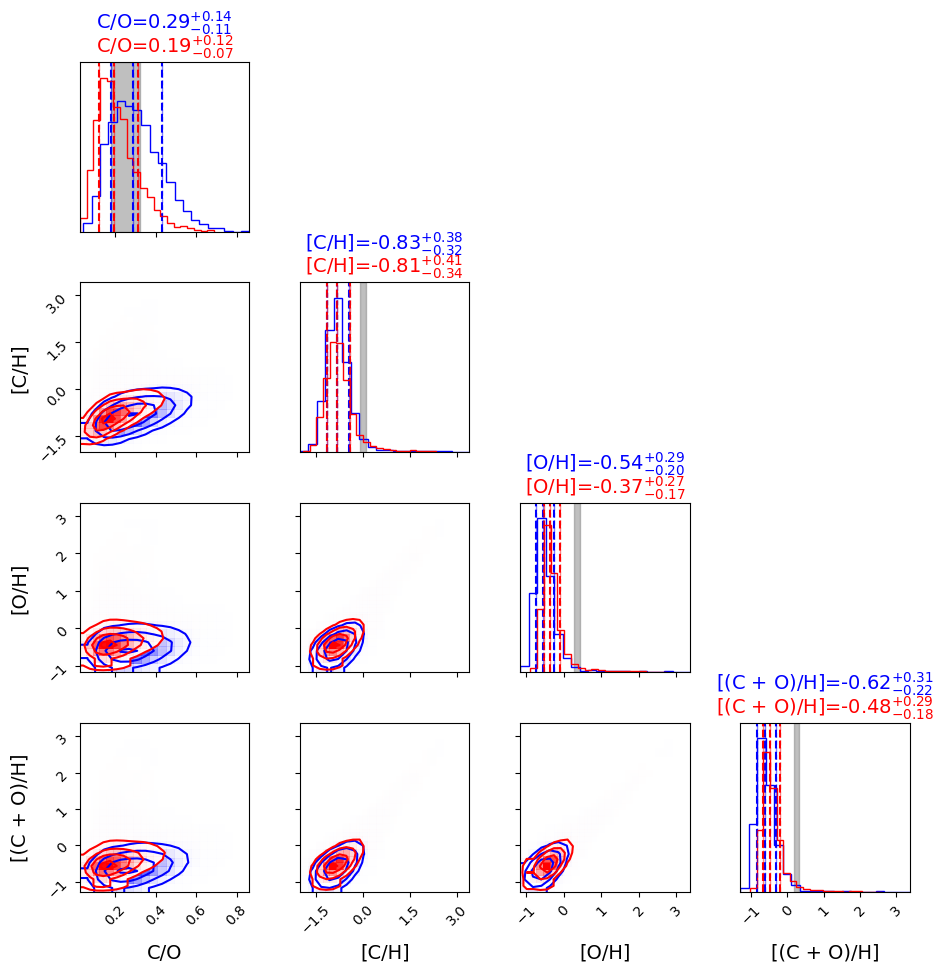}
\caption{Posterior distributions of elemental abundance ratios for free chemistry (blue) and equilibrium chemistry (red) models from the joint fit of IGRINS-2 and \textit{Spitzer}/IRAC. 
Gray shaded regions indicate the stellar 
values taken from \cite{Behmard_2023}. }
\label{fig:elemental_compositions}
\end{figure}

We transform the posterior samples into elemental abundance ratios using Equations~(\ref{eq:C/O_calc})--(\ref{eq:[O/H]_calc}), and display the results from the joint fit with \textit{Spitzer}/IRAC in Figure~\ref{fig:elemental_compositions}.
The free chemistry model constrains $\log(^{12}\rm CO)$, and the derived
elemental ratios are $\rm C/O = 0.29^{+0.14}_{-0.11}$,
$\rm [C/H] = -0.83^{+0.38}_{-0.32}$, $\rm [O/H] = -0.54^{+0.29}_{-0.20}$, and
$\rm [(C+O)/H] = -0.62^{+0.31}_{-0.22}$.\par

The priors and posteriors with EQ model are also summarized in Table~\ref{tab:parameter_prior_posterior}. The corner plot of elemental compositions for the joint fit with IGRINS-2 and \textit{Spitzer}/IRAC is shown in Figure~\ref{fig:elemental_compositions}, and the abundance profiles are shown in Figure~\ref{fig:retrieved_summary}. From the joint fit retrieval with the EQ model, we obtain bounded constraints of
{ $[\rm C/H] = -0.81^{+0.41}_{-0.34}$} and
{ $[\rm O/H] = -0.37^{+0.27}_{-0.17}$}, and derive
{ $\rm C/O = 0.19^{+0.12}_{-0.07}$} and
{ $[(\rm C+O)/\rm H] = -0.48^{+0.29}_{-0.18}$}.
{The chemical abundances from the free and EQ chemistry models agree within $1\sigma$, which is encouraging given their different limitations: free chemistry may not fully account for elemental sequestration, whereas equilibrium chemistry may oversimplify the actual atmospheric conditions.}\par
Furthermore, the elemental abundances retrieved with our EQ model are {consistent within $2\sigma$} with the previous results of \citet{Piskorz_2018}, who obtained $\log z = 1.06^{+0.67}_{-1.17}$ and $\rm C/O = 0.53^{+0.20}_{-0.19}$ from a joint fit of Keck/NIRSPEC and \textit{Spitzer}/IRAC observations.\par
{Our results suggest that the atmosphere of KELT-2A~b may be sub-stellar in
metallicity: the retrieved $[\rm C/H]$ and $[\rm O/H]$ are lower than the corresponding stellar values (C/O $= 0.25 \pm 0.07$,
$[\rm C/H] = 0.01 \pm 0.09$, $[\rm O/H] = 0.35 \pm 0.09$; \citealt{Behmard_2023}),
although the posteriors extend towards higher metallicities and remain
consistent with the stellar values at the $2\,\sigma$ level. The retrieved C/O
ratio is consistent with the stellar value.}\par

\section{Validation and Robustness Test} \label{sec:validation}
\subsection{Injection-Recovery Tests}
\label{subsec:Injection-Recovery}
To evaluate both the quality of the post-eclipse data given the non-detection of a clear planetary signal found in Section~\ref{sec:results}, and the validity of the parameters derived from the retrieval, we perform an injection-recovery test. \par
\begin{figure*}[ht!]
\plotone{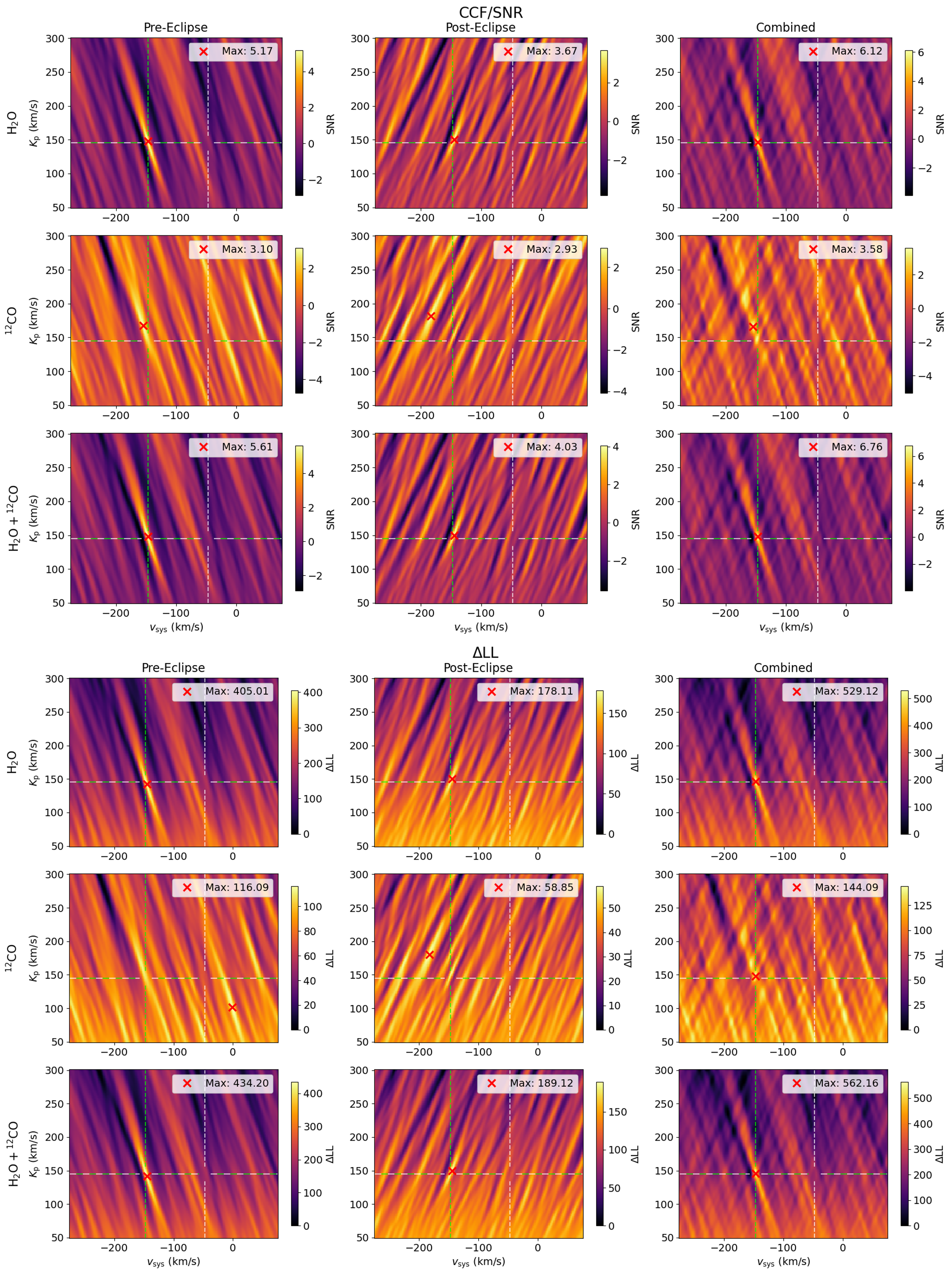}
\caption{
Same as Figures~\ref{fig:SNR_KpVsys_map} and \ref{fig:deltaLL_KpVsys_map}, but for data divided by $1 + F_{\rm scaled}(\lambda, t)$, where $F_{\rm scaled}(\lambda, t)$ is obtained from the retrieved median model, and subsequently re-injected with the same model at a systemic velocity offset by $-100\ \mathrm{km\ s^{-1}}$ from the expected value, keeping $K_{\rm p}$ at its best-fit value. Lime dashed lines indicate the injected velocity and white dashed lines the expected planetary velocity. The top nine panels show the CCF/SNR maps and the bottom nine panels show the $\Delta \ln (\mathcal{L})$ maps.}
\label{fig:injection-recovery}
\end{figure*}

We first remove the planetary signal from the data by dividing the original spectra by $1 + F_{\rm scaled}(\lambda, t)$ computed from the retrieved best-fit model; {if the best-fit model is an accurate representation of the true planetary signal, this division should suppress the planetary feature and cause the peak in the $K_{\rm p}$--$v_{\rm sys}$ map to disappear.}
We then inject the same best-fit model back into the signal-removed data at a systemic velocity offset by $-100\ \mathrm{km\ s^{-1}}$ from the best-fit value and construct $K_{\rm p} - v_{\rm sys}$ maps from the resulting data following the same procedure as in Figures~\ref{fig:SNR_KpVsys_map} and \ref{fig:deltaLL_KpVsys_map}. The results are shown in Figure~\ref{fig:injection-recovery}.

At the position of the real signal, the SNR and $\Delta \ln (\mathcal{L})$ peaks seen in Figures~\ref{fig:SNR_KpVsys_map} and \ref{fig:deltaLL_KpVsys_map} disappear, indicating that these are not merely noise but rather genuine evidence of a planetary signal in the data, and that the best-fit parameters obtained from the retrieval are reasonable given the detected signal. 

At the injected velocity, we successfully recover the signal from the pre-eclipse datasets, demonstrating that these datasets have sufficient quality to detect the planetary signal and the $\rm H_2O$ feature. {For the post-eclipse datasets, the injected $\rm H_2O$ signal is recovered with a CCF/SNR of $\sim 3.7$, about $0.7$ times that of the pre-eclipse datasets ($\sim 5.2$), which is consistent with the ratio expected from the SNR values listed in Table~\ref{tab:observation-summary}. 
The actual post-eclipse data, however, are subject to further factors that would suppress the CCF/SNR beyond this simple scaling: the limited velocity shift of Night 2 during SVD-based telluric removal degrades the planetary signal, and unstable slit guiding and residual spectral artifacts (jaggies) on Night 5 could not be fully removed. Given these additional sources of signal loss, we consider it premature to attribute the post-eclipse non-detection to an astrophysical origin such as spatial asymmetries in the planetary atmosphere. The injection-recovery test also reproduces the absence of a clear $^{12}$CO signal, for which the recovered CCF/SNR remains at a comparably low level in both phases.}

\subsection{Retrieval Robustness Test}
\label{subsec:retrieval_robustness}
\begin{figure*}[ht!]
\plotone{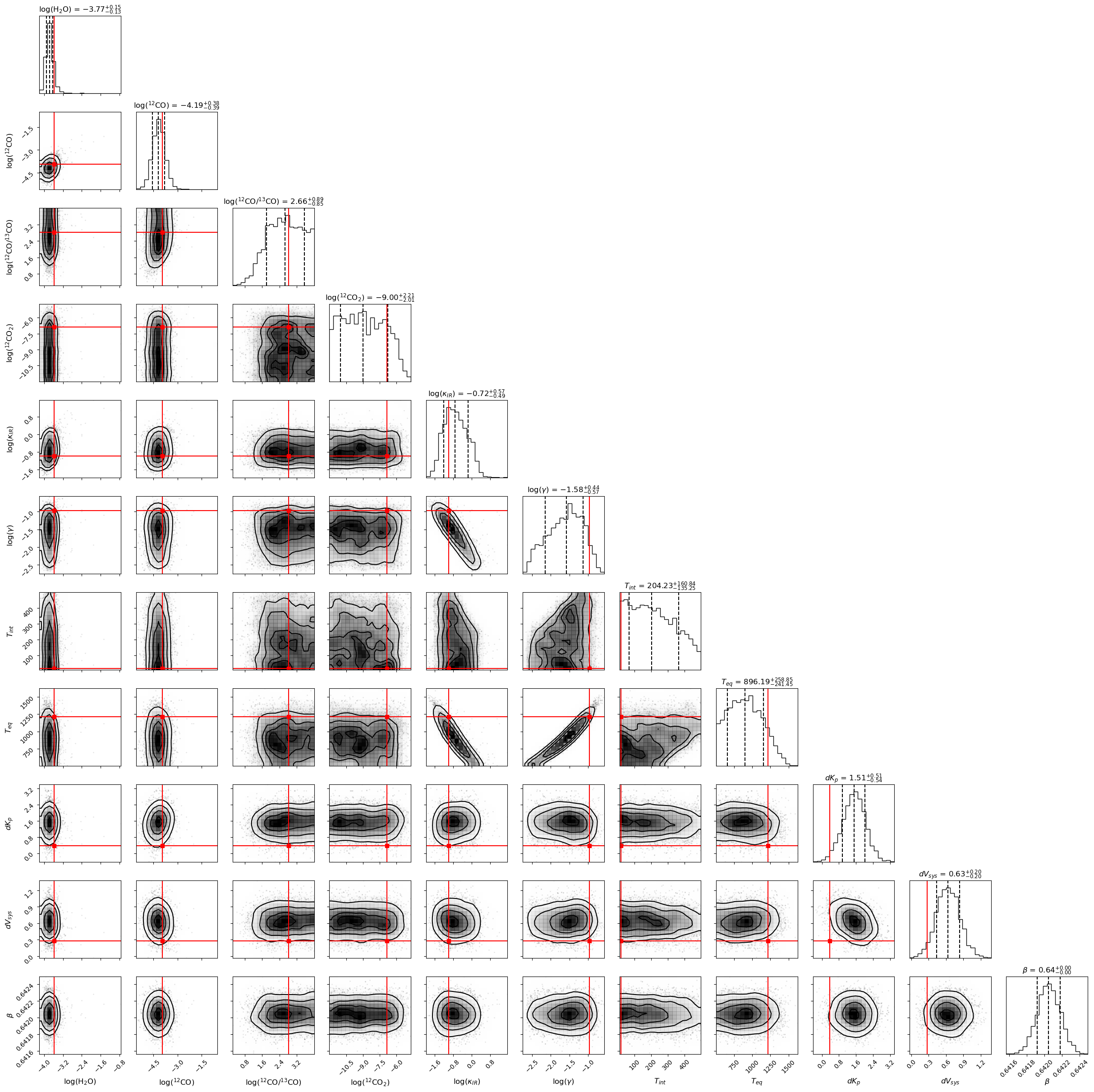}
\caption{Posterior distributions obtained by removing the planetary signal
through division by the best-fit model, re-injecting the same model at a
systemic velocity offset by $-100\ \mathrm{km\ s^{-1}}$ from the expected
value, and performing the retrieval. Red lines indicate the injected
parameter values. The injected parameters are recovered at the $\sim 2\sigma$
level, demonstrating that our retrieval pipeline works correctly.}
\label{fig:re_injected}
\end{figure*}

To validate our retrieval pipeline, we performed a retrieval on the
signal-removed and re-injected data described in
Section~\ref{subsec:Injection-Recovery}. The resulting posterior distributions are shown in Figure~\ref{fig:re_injected}. The retrieved model parameters are consistent with the injected values within $\sim2\sigma$ level, demonstrating that our pipeline works correctly.\par

\begin{figure*}[ht!]
\plotone{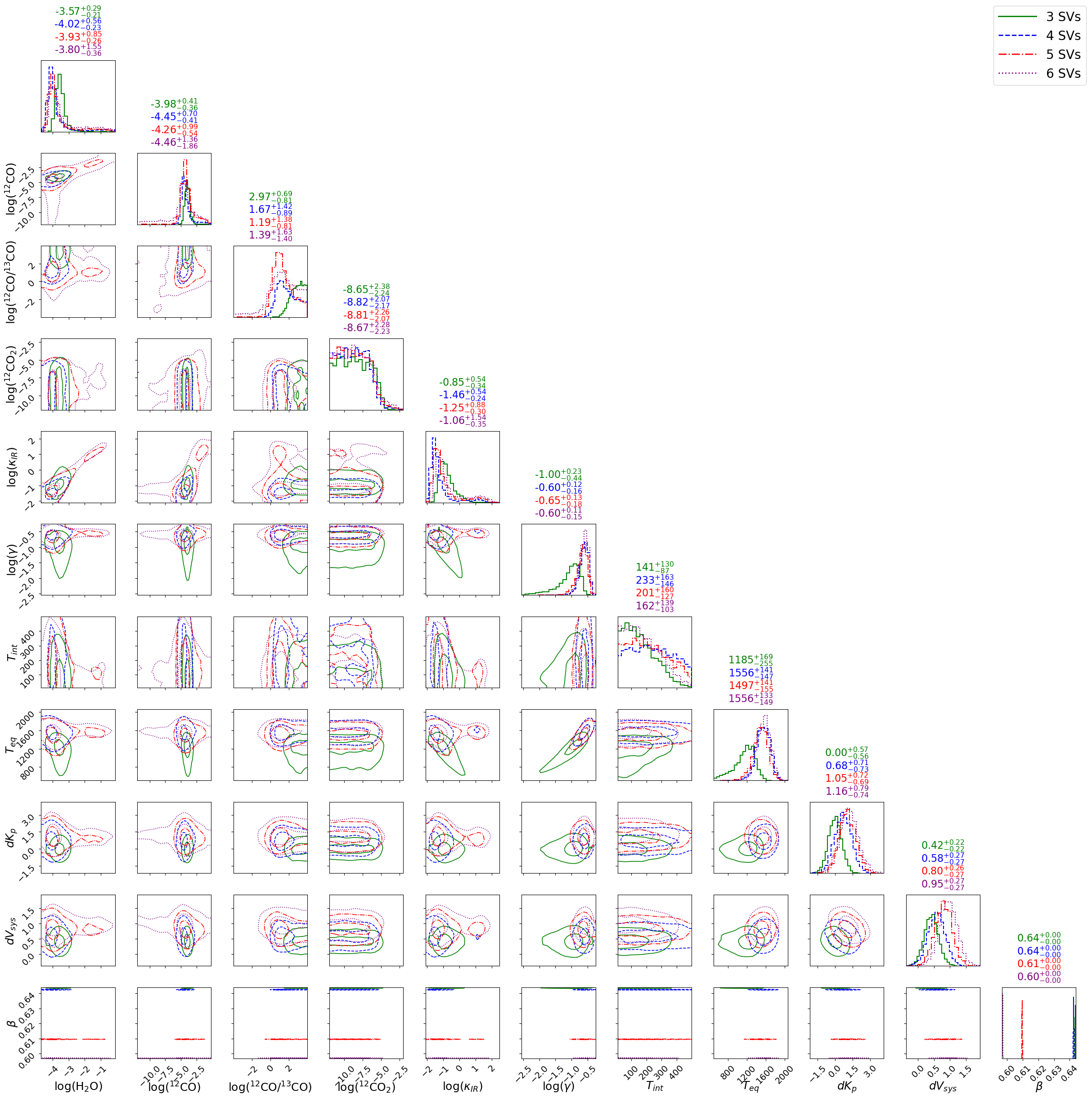}
\caption{Comparison of retrieval results obtained with different numbers of singular values (SVs) removed in the SVD analysis, for the free chemistry model fitted to the IGRINS-2 data alone. Green, blue, red, and purple correspond to 3, 4, 5, and 6 SVs removed for Nights 1–4. For Night 5, one additional SV is removed in each case (i.e., 4, 5, 6, and 7 SVs).}
\label{fig:SVs_comparison}
\end{figure*}

\begin{figure*}[ht!]
\plotone{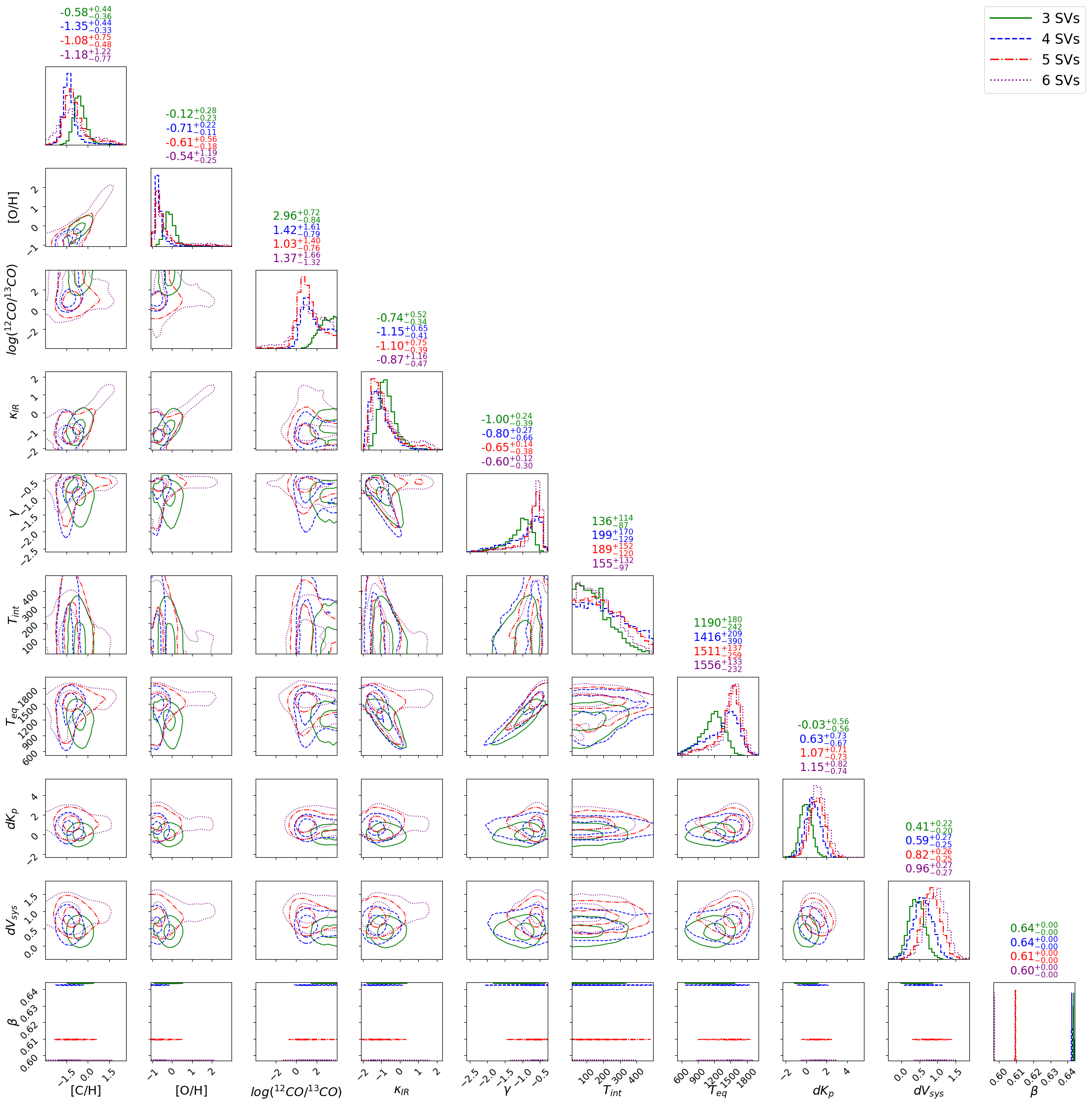}
\caption{Same as Figure \ref{fig:SVs_comparison}, but for the equilibrium chemistry model.}
\label{fig:SVs_comparison_eq}
\end{figure*}

Our retrieval results are robust to the number of singular values used in the
SVD analysis, with retrieved parameters remaining consistent within $1-2\sigma$
(Figure~\ref{fig:SVs_comparison} and Figure~\ref{fig:SVs_comparison_eq}).
{The retrieved uncertainties, however, broaden systematically as more singular values are
removed. This behaviour is expected: as the number of removed components
increases, a growing fraction of the planetary signal is absorbed into the
${\bf U}_k$ of Equation~(\ref{eq:low-order_approx}) and is thus removed together
with the telluric and stellar contamination. The planetary signal retained in
the residual spectra is therefore progressively attenuated, which weakens the
constraints on the retrieved parameters. This effect motivates our choice of the
minimum number of singular values that adequately suppresses the telluric
contamination.}

{The noise scaling factor $\beta$ shows the opposite trend, decreasing
systematically as more components are removed. This is consistent with the
expectation that removing more components captures higher-order systematic
variations in the data (e.g., telluric residuals), leaving less residual
structure that must be compensated for via $\beta$. However, since our noise
model assumes purely white noise, $\beta$ likely underestimates the true noise
level (see Section~\ref{subsec:parameter_estimate}), and this trend does not
necessarily imply that removing more SVD components yields a more accurate noise
treatment.}

{To verify the impact of the Night 2 data (with limited velocity shift) on the overall retrieval results, we also perform retrieval excluding Night 2. We confirm that all parameters agree within $1\, \sigma$.}

\begin{figure*}[ht!]
\plotone{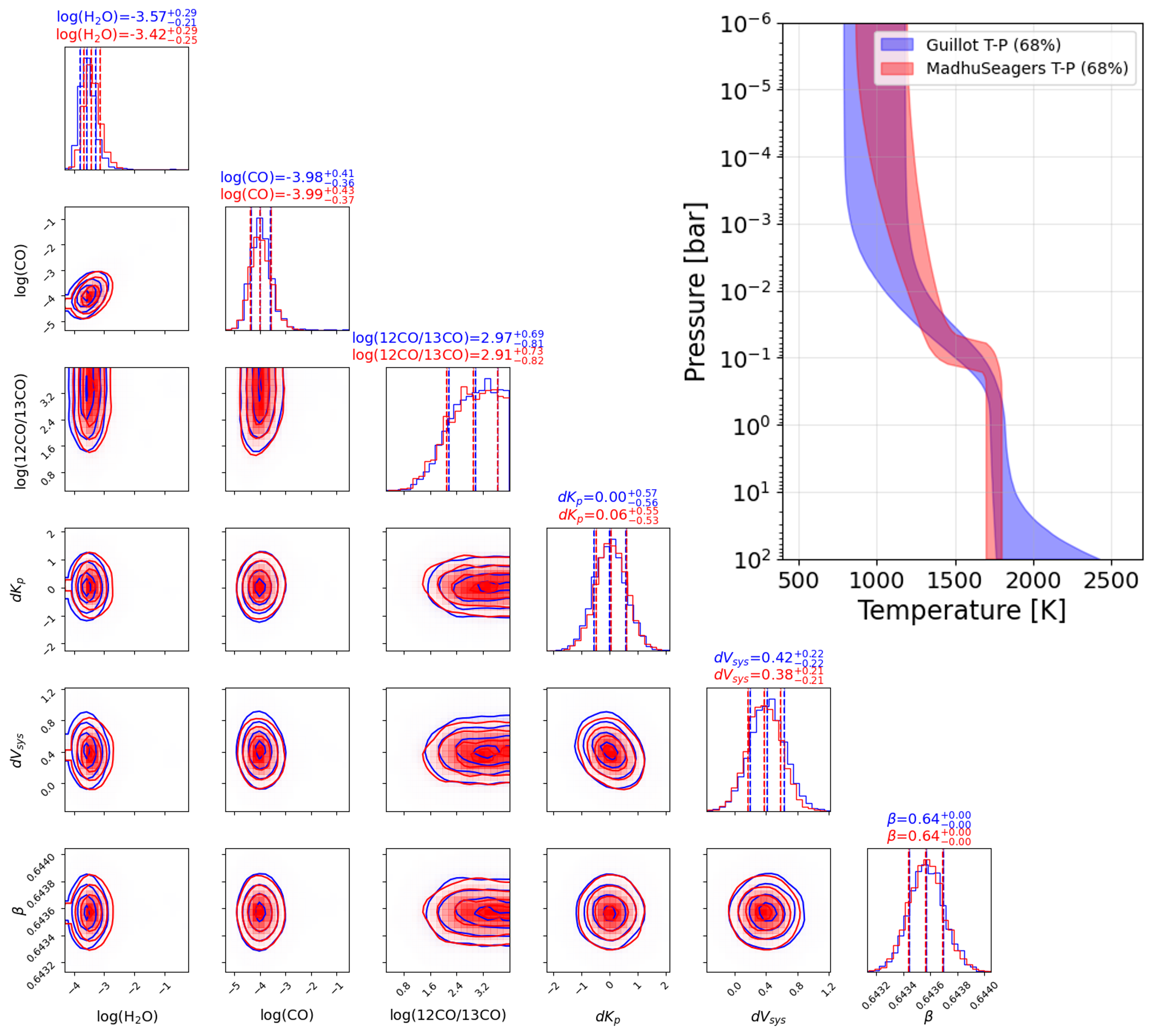}
\caption{Comparison of retrieval results using the temperature profile parameterizations of \citet{Guillot_2010} (blue) and \citet{Madhu_Seager_2009} (red). 
\textit{Left:} Posterior distributions for retrieved parameters except for $T$--$P$ profile parameters. 
\textit{Right:} Retrieved $T$--$P$ profiles with 1$\sigma$ confidence intervals.}
\label{fig:TP_comparison}
\end{figure*}

Furthermore, our results are insensitive to the choice of temperature structure parameterization. We perform a retrieval using the parameterization of \citet{Madhu_Seager_2009} instead of that of \citet{Guillot_2010}. The results are shown in Figure~\ref{fig:TP_comparison}. The two $T$--$P$ profiles agree well, and other parameters also agree within $1\,\sigma$.\par

To quantify which species are statistically supported by the IGRINS-2 data, we compare the Bayesian evidence of the fiducial model with two reduced models: one excluding $^{13}$CO and CO$_2$, which are constrained only as upper limits, and one additionally excluding $^{12}$CO, for which no clear signal is seen in Figure~\ref{fig:SNR_KpVsys_map}. 
We computed the Bayesian log-evidence $\ln\mathcal{Z}$ for each model relative to the full model containing all four species (H$_2$O, $^{12}$CO, $^{13}$CO, and CO$_2$). In all cases, the remaining parameters are consistent within $1\sigma$.
The model excluding CO$_2$ and $^{13}$CO yields the highest $\ln\mathcal{Z}$, but the difference from the fiducial model is ${\Delta\ln\mathcal{Z} = 1.30}$, which is not strong enough to be conclusive according to the classification of \citet{Kass_Raftery_1995}. In contrast, the H$_2$O-only model yields the lowest $\ln\mathcal{Z}$, lower by ${\Delta\ln\mathcal{Z} = 15.75}$ than the model excluding CO$_2$ and $^{13}$CO, which constitutes very strong evidence in favour of including $^{12}$CO in the model. This result reinforces the tentative $^{12}$CO signal seen in Figure~\ref{fig:deltaLL_KpVsys_map}. {We nevertheless do not claim a detection of $^{12}$CO, which requires confirmation by future observations.}

\section{Implication for planet formation}
\label{sec:discussion}
From the joint fit with IGRINS-2 and \textit{Spitzer}/IRAC using the equilibrium chemistry model, the retrieved ${\rm C/O} = {0.19^{+0.12}_{-0.07}}$ {is consistent with the stellar value}, whereas $\rm [C/H]={-0.81^{+0.41}_{-0.34}}$ and $\rm [O/H] = {-0.37^{+0.27}_{-0.17}}$ {are moderately depleted relative to the stellar values, though still consistent at the $\sim2\sigma$ level. Moreover, as shown in Section~\ref{subsec:retrieval_robustness}, the uncertainties broaden as more singular values are removed, and the posteriors then accommodate metallicities that are stellar or even super-stellar. We therefore regard the sub-stellar metallicities inferred here as suggestive rather than conclusive.} Future observations with JWST, \textit{Ariel} or Keck/HISPEC will be necessary to confirm these values. \par 
{If the median values are confirmed, the composition of this planet can be explained by gas accretion outside the CO snowline with limited planetesimal accretion. Between the snowlines of the major C, O-bearing species ($\rm H_2O$, $\rm CO_2$, $\rm CO$), a sub-stellar metallicity is expected but with a super-stellar C/O ratio \citep{Oberg_2011, Madhusudhan_2014, Notsu_2020, Khorshid_2023}, which is inconsistent with our stellar C/O. Outside the CO snowline, in contrast, both C and O are almost fully condensed, resulting in metal-depleted gas with a near-stellar C/O ratio. Planetesimal accretion would raise the metallicity and lower the C/O below the stellar value \citep{Mordasini_2016, Cridland_2019}, and is therefore limited in this planet.}
{We further suggest that this planet may have undergone disk-free migration. The CO snowline is located at several tens of au \citep{Qi_2013}, and migration from such a distance would involve substantial planetesimal capture along the way, resulting in a high metallicity \citep{Shibata_2020, Shibata_2022}, which is inconsistent with our measurement. Eccentricity excitation by the Kozai-Lidov mechanism or planet-planet scattering, followed by tidal dissipation at periastron, has been proposed as a formation pathway for hot Jupiters \citep{Dawson_Johnson_2018}, and is consistent with our atmospheric measurements. Future obliquity measurements would help test this scenario \citep{Narita_2009, Winn_2010}, although tidal realignment may erase the signature for cool host stars \citep{Albrecht_2012}.}

We find a $^{12}$CO/$^{13}$CO ratio of 
${>21}$ ($97.5\%$ lower limit) {from the joint fit with IGRINS-2 and \textit{Spitzer}/IRAC with the equilibrium chemistry model}, consistent with values typical of the ISM and Solar System, but also consistent with the $^{13}$C-enriched isotope ratio found in CO gas in a protoplanetary disk TW Hya ($^{12}$C/$^{13}$C $=21\pm5$; \citealt{Yoshida_2022}). 
Future observations with improved precision on the isotope ratio, combined with comparisons between protoplanetary disk observations and models, may provide a new diagnostic for planet formation environment \citep[e.g.,][]{Bergin_2024, Lee_2024}.\par
We emphasize that significant uncertainties remain in linking observed atmospheric compositions to planet formation scenarios. These include uncertainties in disk structure, physical properties, chemical evolution, and the relationship between atmospheric and bulk compositions (e.g., \citealt{Schneider_Bitsch_2021, Molliere_2022, Notsu_2022, Penzlin_2024, Williams_2025, Ohno_2026}). \par

\section{Summary}
We have presented a high-resolution emission spectroscopic analysis of the hot Jupiter KELT-2A~b using IGRINS-2 on Gemini-North. 
We detect $\rm H_2O$ in the dayside atmosphere at SNR = {6.57}, {validated by injection-recovery tests.} While $\rm ^{12}CO$ is identified in the log-likelihood map near the expected planetary velocity {, it is not definitively detected.}
From the retrieval analysis, the temperature–pressure profile exhibits a non-inverted structure around ${\sim 0.1\ \rm bar}$, in agreement with the earlier Keck/NIRSPEC and \textit{Spitzer} analysis of \citet{Piskorz_2018} and consistent with other hot Jupiters at comparable equilibrium temperatures. 
We obtain bounded constraints of $\rm log(H_2O) = {-3.57^{+0.29}_{-0.21}}$ and $\rm log(^{12}CO) = {-3.98^{+0.41}_{-0.36}}$ {with the free chemistry model from IGRINS-2 only retrieval}. The inclusion of \textit{Spitzer}/IRAC photometry data points yields tighter constraints of $\rm log(H_2O) = {-3.78^{+0.27}_{-0.20}}$ and $\rm log(^{12}CO) = {-4.17^{+0.39}_{-0.32}}$ {with the free chemistry model}. 
The free-chemistry and equilibrium-chemistry retrievals agree within $1\,\sigma$. The derived elemental abundance ratios {from the joint fit of IGRINS-2 and \textit{Spitzer}/IRAC with the EQ chemistry model}, ${\rm C/O = 0.19^{+0.12}_{-0.07}}$, ${\rm [C/H]=-0.81^{+0.41}_{-0.34}}$, and ${\rm [O/H] = -0.37^{+0.27}_{-0.17}}${, indicate a stellar C/O ratio together with a moderately sub-stellar metallicity ($\rm C/O = 0.25 \pm 0.07$, $\rm [C/H] = 0.01 \pm 0.09$, and $\rm [O/H] = 0.35 \pm 0.09$ for the host star), although both remain consistent with the host-star values at the $\sim2\sigma$ level. If these median values are confirmed, they are difficult to explain by gas accretion between the $\rm H_2O$ and CO snowlines, which predicts a super-stellar C/O ratio, and instead favour the accretion of metal-depleted gas outside the CO snowline with limited planetesimal accretion. Such a composition disfavours long-range disk migration, which would entail substantial planetesimal capture, and is therefore consistent with a disk-free, high-eccentricity migration pathway.}
Our retrieval results are robust to the number of SVs removed in the SVD analysis and to the choice of temperature profile parameterization, with all retrieved atmospheric parameters consistent within $1-2\sigma$ across these tests{, although the retrieved uncertainties broaden as more singular values are removed. Higher-precision abundance measurements from future observations will be needed to confirm these values and to place the planet firmly within a formation scenario.}
\label{sec:summary}

\begin{acknowledgments}
We would like to thank Hajime Kawahara and Kento Masuda for the fruitful discussions on correlated noise.
This research is based on data obtained at the international Gemini Observatory, a program of NSF NOIRLab, via the time exchange program between Gemini and the Subaru Telescope. The international Gemini Observatory at NOIRLab is managed by the Association of Universities for Research in Astronomy (AURA) under a cooperative agreement with the U.S. National Science Foundation on behalf of the Gemini partnership: the U.S. National Science Foundation (United States), National Research Council (Canada), Agencia Nacional de Investigaci\'{o}n y Desarrollo (Chile), Ministerio de Ciencia, Tecnolog\'{i}a e Innovaci\'{o}n (Argentina), Minist\'{e}rio da Ci\^{e}ncia, Tecnologia, Inova\c{c}\~{o}es e Comunica\c{c}\~{o}es (Brazil), and Korea Astronomy and Space Science Institute (Republic of Korea). This work was enabled by observations made from the Gemini North telescope, located within the Maunakea Science Reserve and adjacent to the summit of Maunakea. We are grateful for the privilege of observing the Universe from a place that is unique in both its astronomical quality and its cultural significance. This work is partly supported by JST SPRING, Grant Number JPMJSP2108, JSPS KAKENHI Grant Numbers JP24H00017, JP25K24620,  JP26H01402, JP24K00689, JP23K13155, JP24K00674, JP23H05441, JP26K00755, and JSPS Fellows grant Numbers JP23KJ0329, JP25KJ1040, and the Start-up Research Grant as one of the University of Tokyo Excellent Young Researchers 2024.
We acknowledge financial support from the Agencia Estatal de Investigaci\'on of the Ministerio de Ciencia e Innovaci\'on MCIN/AEI/10.13039/501100011033 and the ERDF “A way of making Europe” through projects PID2021-125627OB-C32 and PID2024-158486OB-C32. This work is supported by the European Union (ERC AdvG SPEAR, GA 101200674). Views and opinions expressed are however those of the authors only and do not necessarily reflect those of the European Union or the European Research Council. Neither the European Union nor the granting authority can be held responsible for them.
T.C.Y. acknowledges support from the European Union (ERC Starting Grant DiscEvol, project number 101039651) and from Fondazione Cariplo, grant No. 2022-1217. Views and opinions expressed are, however, those of the author(s) only and do not necessarily reflect those of the European Union or the European Research Council. Neither the European Union nor the granting authority can be held responsible for them.
\end{acknowledgments}

{
The reduced datasets used in this work are publicly available on 
Zenodo\footnote{\url{https://doi.org/10.5281/zenodo.20908908}}.
}





%
\facilities{Gemini-North (IGRINS-2), \textit{Spitzer} (IRAC)}

\software{
    astropy \citep{astropy},
    corner \citep{corner},
    easyCHEM \citep{Lei_Molliere_2025},
    HELIOS-K \citep{Grimm_2015, Grimm_2021},
    IGRINS-2 PLP \citep{IGRINS-2_PLP}
    Matplotlib \citep{matplotlib},
    NumPy \citep{Numpy},
    petitRADTRANS \citep{Molliere_2019, Blain_2024},
    PyAstronomy \citep{pya},  
    PyMultiNest \citep{Buchner_2014}, 
    Scipy \citep{SciPy}
    }



\bibliography{sample701}{}
\bibliographystyle{aasjournalv7}



\end{document}